\documentclass[a4paper,11pt]{article}
\usepackage{jheppub}
\usepackage{bm}
\usepackage{pgfplots}
\pgfplotsset{compat=newest}
\usetikzlibrary{shadows.blur}
\usepackage{float}
\def\beq{\begin{equation}}
\def\eeq{\end{equation}}
\def\bea{\begin{eqnarray}}
\def\eea{\end{eqnarray}}

\renewcommand{\v}[1]{ \ensuremath{ {\bm{#1}} }}                  
\parskip=\itemsep               %?

\title{Born-Oppenheimer Renormalization group for High Energy Scattering: CSS, DGLAP and all that.}
\author[a]{Haowu Duan,}
\author[a,b]{Alex Kovner}
\author[c]{and Michael Lublinsky}

\affiliation[a]{Physics Department, University of Connecticut, 2152 Hillside Road, Storrs, CT 06269, USA}
\affiliation[b]{ExtreMe Matter Institute EMMI,
GSI Helmholtzzentrum fuer Schwerionenforschung GmbH,
Planckstrasse 1,
64291 Darmstadt,
Germany}
\affiliation[c]{Physics Department, Ben-Gurion University of the Negev, Beer Sheva 84105, Israel}

\abstract{In \cite{one}, we have introduced the Born-Oppenheimer (BO) renormalization group approach to high energy 
hadronic collisions and derived the BO approximation for the light cone wave function of a fast moving projectile hadron.
In this second paper, we utilize this wave function 
to derive the BO evolution of partonic distributions in the hadron  -- the gluon transverse momentum 
and integrated parton distributions  (TMD  and PDF respectively).
The evolution equation for the TMD contains a linear and a nonlinear term. The linear term reproduces the Collins-Soper-Sterman (CSS) equation with a physical relation between the transverse and longitudinal resolution scales. We explain how this equivalence arises, even though the BO and CSS
cascades are  somewhat different in structures. 
The nonlinear  term in the evolution has a very appealing physical meaning: it is a correction due to
stimulated emission,  which enhances emission of gluons (bosons) into states with a nonzero occupation.
 For the evolution of the PDF we again find a linear and  nonlinear term. At not very small Bjorken $x$, the linear term recovers the DGLAP equation in the leading logarithmic approximation. At small  $x$ however there are contributions from gluon splittings which are in the BFKL kinematics leading to a modification of the DGLAP equation. The nonlinear terms have the same physical origin as in the equation for the TMD -- the stimulated emission corrections. Interestingly the nonlinear corrections are the most important for the virtual terms, so that the net correction to the DGLAP is negative and mimics shadowing, although the physical origin of the nonlinearity is very different.}

\begin{document}
\maketitle

\pagestyle{empty}

\mbox{}

\pagestyle{plain}

\setcounter{page}{1}
\author{}

\abstract{ }
\dedicated{}
\preprint{}

\date{\today}
%\begin{document}

%%%%%%%%%%%%%%%%%%%%%%%%%%%%%%%%%%%%%%%%%%%%%%%%%%%%%%%%%%%%%%%%%%%%%%
\section{Introduction }

\begin{figure}[H] % Use figure environment for captions
\centering % Center the plot
\begin{tikzpicture}
\begin{axis}[scale=1.2,
        axis lines=left,             % Keep axes at the left and bottom
        axis line style={-latex},
	 xmin=0,   xmax=4,
	 ymin=0,   ymax=2.5,
         %extra x ticks={-1,1},
	 %extra y ticks={-2,2},
	 xtick=\empty,
	 ytick=\empty,
	 xlabel={$k^-$}, % Optionally keep labels without numbers
         ylabel={$\Psi$},
          x label style={at={(axis description cs:1,0)}}, 
          y label style={rotate=270, at={(axis description cs:0,1)}},
	%extra tick style={grid=major}, 
	clip=false,
	]    
       	%\addplot[dash pattern=on 5pt off 3pt] coordinates {(0,2) (4,2)};
	 \addplot[domain=0:2.5, dash pattern=on 5pt off 3pt] ({2},{x}); % A vertical line at x=2
	 \addplot[domain=0:2.5, dash pattern=on 5pt off 3pt] ({2.8},{x});
	 \node at (axis description cs:0.5,-0.1) [anchor=south] {$E$};
	 \node at (axis description cs:0.7,-0.1) [anchor=south] {$Ee^\Delta$};
	 % \node at (axis description cs:-0.15,0.25) [anchor=west] {$X_{Bj}$}; 
	  % Here we add the brackets and labels
	   \addplot[domain=0:2.8, smooth, thick, blue, mark=none] {2*(x/2.8)^2 * (2.8-x) * sin(pi*x/2.8)*500 / (1+x)^2} 
    node[pos=0.4, above, font=\small, xshift=-20pt] {\textcolor{blue}{Projectile}};
    %---------------------------------------------------------------------------------------------------
       %\addplot[domain=2.8:4, smooth, thick, red, mark=none] {2 * (1 - (x - 2.8) / (4 - 2.8))^2 } 
       %     node[pos=0.2, left, yshift=5pt, font=\small, xshift=35pt] {\textcolor{red}{Target}};
 \addplot[domain=2.8:4, smooth, thick, red, mark=none] {(5)*exp(-20*(x-2.8))*(20*x-56)} 
    node[pos=0.2, left, yshift=5pt, font=\small, xshift=50pt] {\textcolor{red}{Target}};    % \addplot[domain=0:2.8, smooth, thick, blue] {2*(x/2.8)^2 * (2.8-x) * sin(pi*x/2.8)*500 / (1+x)^2};
      {\color{red}
    \draw[decorate,decoration={brace,amplitude=5pt,mirror},very thick] (0,0) -- (2,0) node[midway,below=8pt,font=\small] {Slow/ Valence};
    \draw[decorate,decoration={brace,amplitude=5pt,mirror},very thick] (2,0) -- (2.8,0) node[midway,below=8pt,font=\small] {Fast};
    }
	\end{axis}
\end{tikzpicture}
\caption{A qualitative picture of the projectile and target wave functions in the $k^-$ ordering scheme.}
\label{Fig} % Optional: label for referencing the figure
\end{figure}
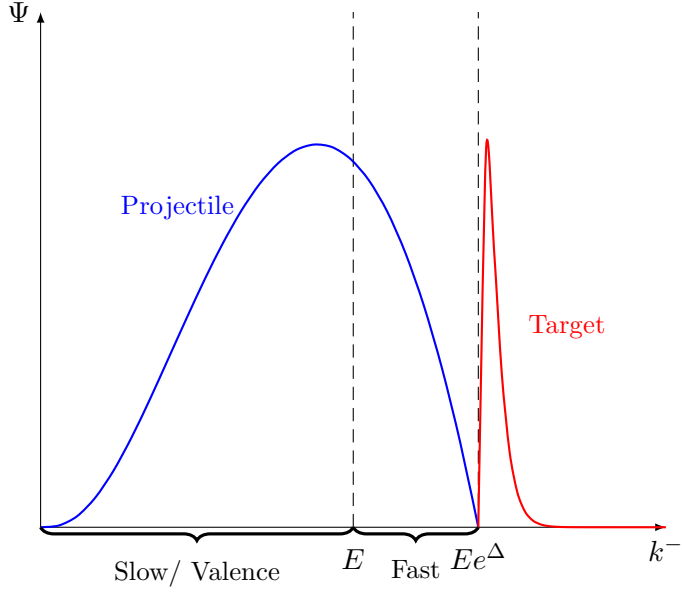

In the previous paper \cite{one} we have developed the Born-Oppenheimer renormalization group approach to QCD evolution of hadronic observables. It is based on
frequency factorization between projectile ($P$) and target ($T$) hadrons, see Fig. \ref{Fig}.  %The $S$-matrix operator $\hat S$ is a quantum operator on the Hilbert space which is a product of the Hilbert spaces of the projectile  and the target. 
%The projectile is taken to have the total longitudinal momentum $P^+$ while the central mass collision energy squared is $s$.

The projectile light cone wave function (LCWF)
$|\Psi_P\rangle$ is calculated using a sequence of  Born-Oppenheimer approximations.   The gluon field modes in the projectile are split, on the basis of their frequency $k^-$,  between "slow" (valence) acting as a frozen background for the "fast" modes, Fig. \ref{Fig}.  Increasing the frequency cutoff that separates the modes, adds more "active" modes to the wave function leading to the evolution of physical observables, such as the scatterning amplitude. %the wave function, which we  refer to  as the Born Oppenheimer renormalization group.

In \cite{one} we have derived  the  LCWF of the projectile evolved from the initial frequency scale 
$E_0$ to the final scale $E$ determined by the frequency of the dominant modes in the target.  
The physical meaning of the frequency scale $E$ depends on the process/observable. For example for the calculation of the evolution of a cross section with energy, $E$ stands for the inverse of the Ioffe time cutoff.
 For DIS where the frequency is increased due to increase in $Q^2$, the scale $E$ is naturally related to the transverse resolution scale (much more on this later). 
 
 In the projectile light cone gauge $A^+=0$, the LCWF of  \cite{one} reads
\begin{equation}\label{psiel1}
 |\Psi_P\rangle_E={\cal P}\exp\left\{i\int_{E_0}^{E}d p^- {\cal G}(p^-)\right\}|\Psi_P\rangle_{E_0}
 %|\Psi_E\rangle={\cal P}\exp\{i\int_{E_0}^{E}d p^- G(p^-)\}|\Psi_{E_0}\rangle
 \end{equation}
where ${\cal P}$ denotes "path ordering" from low to high frequency, which we will sometimes refer to as rapidity, and 
\begin{eqnarray}
%G(p^-)&\equiv& \int \frac{dp^+d^2p}{(2\pi)^3}\ \delta(p^--\frac{p^2}{p^+})G(p^+,p^2); \nonumber\\
{\cal G}(p^-)&\equiv& \int_p %\frac{dp^+d^2p}{(2\pi)^3}\ 
\delta(p^--\frac{\v p^2}{2p^+})G(p^+,\v p^2); 
 \end{eqnarray}
 with $G(p^+,\v p)$ given by  
 \begin{equation}
    \begin{split}
%    G(p^+,\v p)&=g \int_{k^-<p^-; \ (k-p)^-<p^-}\frac{dk^+}{2\pi} \frac{d^2\v k}{(2\pi)^2}  A^a_{i}(k^+,k)\frac{2p^+(k^+-p^+)}{k^+}\, \\
%&\times\left\{\left[\delta_{ki}\delta_{jl}\left(\frac{2k^+}{p^+}-1\right)+\epsilon_{ki}\epsilon_{jl}\right]\frac{\v p_j}{\v p^2} A^{\dagger }_{l}(p^+,p)T^a A^{\dagger }_{k}(k^+-p^+,k-p)
%\right. \\
%&-\left.\left[\delta_{ki}\delta_{jl}\left(\frac{2k^+}{k^+-p^+}-1\right)+\epsilon_{ki}\epsilon_{jl}\right]\frac{\v p_j}{\v p^2} A^{\dagger }_{l}(k^+-p^+,k-p)T^a A^{\dagger}_{k}(p^+,p)\right\} \\
%&
%+h.c.
   G(p^+, \v p)&=g \int_{k^-<p^-; \ (k-p)^-<p^-}%\frac{dk^+}{2\pi} \frac{d^2  \v k}{(2\pi)^2}  
    A^a_{i}(k^+,\v k)\frac{2p^+(k^+-p^+)}{k^+}\, \\
&\times\left\{\left[\delta_{ki}\delta_{jl}\left(\frac{2k^+}{p^+}-1\right)+\epsilon_{ki}\epsilon_{jl}\right]\frac{\v  p_j}{ \v p^2} A^{\dagger }_{l}(p^+,\v p)T^a A^{\dagger }_{k}(k^+-p^+,\v k-\v p)
\right. \\
&-\left.\left[\delta_{ki}\delta_{jl}\left(\frac{2k^+}{k^+-p^+}-1\right)+\epsilon_{ki}\epsilon_{jl}\right]\frac{\v p_j}{\v p^2} A^{\dagger }_{l}(k^+-p^+,\v k-\v p)T^a A^{\dagger}_{k}(p^+,\v p)\right\} 
+h.c.
\end{split}
\end{equation}
Here $A$ are the gluonic fields (not creation and/or annihilation operators) normalized as
\begin{equation}\label{A}
\langle 0| A(p)A^\dagger(p)|0 \rangle=\frac{(2\pi)^3}{2p^+}
\end{equation}
A representation that we will find convenient is
\begin{equation}\label{ca}
\begin{split}
%G(p^+,\v p)= A^{\dagger }_{i}(p^+,\v p)C_i(p^+,\v p)+  A_{i}(p^+,\v p)C_i^\dagger(p^+,\v p)
G(p^+, \v p)= A^{\dagger }_{i}(p^+,  \v p)C_i(p^+,  \v p)+  A_{i}(p^+,  \v p)C_i^\dagger(p^+,  \v p)
\end{split}
\end{equation}
with  
\begin{equation}\label{c}
\begin{split}
%C_i^{a}(p^+, \v p)
%=&g \int_{p^->k^-, p^->(k-p)^-} \frac{dk^+d^2\v k}{(2\pi)^3}  \frac{2p^+(k^+-p^+)}{k^+}\, \\
%&\times\left\{\left[\delta_{kl}\delta_{ji}\left(\frac{2k^+}{p^+}-1\right)+\epsilon_{lk}\epsilon_{ji}\right]+\left[\delta_{ki}\delta_{jl}\left(\frac{2k^+}{k^+-p^+}-1\right)+\epsilon_{ik}\epsilon_{jl}\right]\right\} \\
%&\times \frac{\v p_j}{\v p^2} A^{\dagger}_{l}(k^+-p^+,\v k-\v p)T^a  A_{k}(k^+,\v k)
C_i^{a}(p^+,   \v p)
=&g \int_{p^->k^-, p^->(k-p)^-} %\frac{dk^+d^2  k}{(2\pi)^3}
  \frac{2p^+(k^+-p^+)}{k^+}\, \\
&\times\left\{\left[\delta_{kl}\delta_{ji}\left(\frac{2k^+}{p^+}-1\right)+\epsilon_{lk}\epsilon_{ji}\right]+\left[\delta_{ki}\delta_{jl}\left(\frac{2k^+}{k^+-p^+}-1\right)+\epsilon_{ik}\epsilon_{jl}\right]\right\} \\
&\times \frac{ \v p_j}{ \v p^2} A^{\dagger}_{l}(k^+-p^+, \v k- \v p)T^a  A_{k}(k^+, \v k)
\end{split}
\end{equation}
In the above expressions it is implied that the momentum $p$ corresponds to the highest frequency mode in the interaction vertex, i.e. $p^->max\{k^-, (p-k)^-\}$.

In the BFKL regime \cite{bfkl1,bfkl2,bfkl3}, i.e. assuming $p^+\ll k^+$, the operator $C$ reduces to the "classical field" produced by valence charges which is routinely used in discussions of high energy evolution. In this paper we are however not interested in this limit and will use the full form given in \eqref{c}. %In  the DGLAP regime, for $|\v p|^2\gg |\v k|^2$ and $k^+\sim p^+$ it contains one slow and one fast mode but we still find this form useful in calculations.

As a shorthand, we introduce
\begin{equation}\label{F}
\begin{split}
%F_{lk}^i(k,p)&=\frac{2p^+(k^+-p^+)}{k^+}\left\{\left[\delta_{kl}\delta_{ji}\left(\frac{2k^+}{p^+}-1\right)+\epsilon_{lk}\epsilon_{ji}\right]+\left[\delta_{ki}\delta_{jl}\left(\frac{2k^+}{k^+-p^+}-1\right)+\epsilon_{ik}\epsilon_{jl}\right]\right\} \frac{\v p_j}{\v p^2}
F_{lk}^i(k,p)&=\frac{2p^+(k^+-p^+)}{k^+}\left\{\left[\delta_{kl}\delta_{ji}\left(\frac{2k^+}{p^+}-1\right)+\epsilon_{lk}\epsilon_{ji}\right]+\left[\delta_{ki}\delta_{jl}\left(\frac{2k^+}{k^+-p^+}-1\right)+\epsilon_{ik}\epsilon_{jl}\right]\right\} \frac{ \v p_j}{\v  p^2} \\
&=\frac{4p^+(k^+-p^+)}{k^+}\left\{\delta_{kl}\delta_{ji}\frac{k^+}{p^+}+\delta_{ki}\delta_{jl}\frac{k^+}{k^+-p^+}-\delta_{kj}\delta_{il}\right\} \frac{\v p_j}{\v p^2} 
\end{split}
\end{equation}
such that
\begin{equation}
\begin{split}
 %C_i^{a}(p^+,\v p)&=g\int_{k^-<p^-;\ \ (k-p)^-<p^-}\frac{dk^+}{2\pi} \frac{d^2\v k}{(2\pi)^2} F^i_{lk}(k,p) A^{\dagger}_{l}(k^+-p^+,\v k-\v p)T^a  A_{k}(k^+,\v k)
 C_i^{a}(p^+, \v p)&=g\int_{k^-<p^-;\ \ (k-p)^-<p^-}%\frac{dk^+}{2\pi} \frac{d^2  k}{(2\pi)^2}
  F^i_{lk}(k,p) A^{\dagger}_{l}(k^+-p^+, \v k-\v  p)T^a  A_{k}(k^+,  \v k)
\end{split}
\end{equation}
\begin{equation}
\begin{split}
 %C_i^{a\dagger}(p^+,\v p)&=g\int_{k^-<p^-;\ \ (k+p)^-<p^-}\frac{dk^+}{2\pi} \frac{d^2\v k}{(2\pi)^2} F^i_{kl}(k+p,p) A^{\dagger}_{l}(k^++p^+,\v k+\v p)T^a  A_{k}(k^+,\v k)
 C_i^{a\dagger}(p^+, \v p)&=g\int_{k^-<p^-;\ \ (k+p)^-<p^-}%\frac{dk^+}{2\pi} \frac{d^2  k}{(2\pi)^2} 
 F^i_{kl}(k+p,p) A^{\dagger}_{l}(k^++p^+,  \v k+ \v p)T^a  A_{k}(k^+,\v  k)
\end{split}
\end{equation}

In \cite{one} we have also derived the expression for the forward $S$-matrix in the dilute target limit. 
It is given by \footnote{We use the shorthand notation $\int_k\equiv \int \frac{d^2\v kdk^+}{(2\pi)^3}$.}
  \begin{equation}
 %S=1-\int_q\langle a^{\dagger b}_j(q)a^b_j(q)\rangle O(q)
  S=1\,-\,{1\over 2(N_c^2-1)}\,\int_q %\frac{dq^-d^2  q}{(2\pi)^3}\, 
 \langle\Psi_T| \gamma^{\dagger b}_j(q)\gamma^b_j(q)|\Psi_T\rangle\,
 O(q)
 \label{S}
 \end{equation}
 $\langle\Psi_T|\gamma^{\dagger b}_j(q)\gamma^b_j(q)|\Psi_T\rangle$ is the average of the targets fields (denoted as $\gamma_j^a$)   over the target wave function (ensemble of fields), 
 and  $O$ is the matrix element in the projectile Hilbert space
 %where $\langle a^{\dagger b}_j(q)a^b_j(q)\rangle$ 
 %denotes the average of the target fields over the target wave function, and $O$ is the average over  the projectile Hilbert space
 \begin{eqnarray}\label{scat}
%O(q)&=&\langle {\cal C}_i^{a} (q^+,q){\cal C}^{a\dagger}_i(q^+,q)\rangle;\\
%{\cal C}_i^{a\dagger}(q^+,q)&=&g\int_{k^-<q^-}\frac{dk^+}{2\pi} \frac{d^2\v k}{(2\pi)^2} f^i_{lk}(k,q) A^{\dagger}_{l}(k^++q^+,\v k+\v q)T^a  A_{k}(k^+,\v k)
&&O(q)=\langle \Psi_P |{\cal C}_i^{a} (q^-,\v q){\cal C}^{a\dagger}_i(q^-,\v q)|\Psi_P\rangle_E;\\ \nonumber \\
&&{\cal C}_i^{a\dagger}(q^-,\v q)=g\int_{k:k^-<q^-}
%\frac{dk^+}{2\pi} \frac{d^2  k}{(2\pi)^2} 
f^i_{lk}(k,q) A^{\dagger}_{l}(k^+,  \v k+ \v q)T^a  A_{k}(k^+,\v  k)
\end{eqnarray}
with $f^i_{jk}$ given by 
\begin{equation}\label{fijk}
 f^i_{jk}(k,q)=2k^+\frac{\v q^i}{\v q^2}\delta^{jk} +4k^+\epsilon^{il}\frac{\v q^l}{\v q^2}\epsilon^{jk} \frac{1}{\frac{2q^-k^+}{\v q^2}-1}
 \end{equation}
In the above expressions $q^-\simeq E$ is the frequency of the target fields (the  distribution of frequencies in the target is assumed to be sharply peaked around $E$), and is the final frequency for the evolution of the projectile.  The transverse momentum $\v q$  of the target field  is also the transverse momentum  transfer in the scattering.

The scattering amplitude  \eqref{scat} physically contains two distinct contributions
\begin{equation}
O(q)=O_1(q)+O_2(q)
\end{equation}
with 
\begin{equation}\label{1ps}
%O_1(q)=g^2N_c \int_{k^-<q^-}  \frac{dk^+}{2\pi} \frac{d^2\v k}{(2\pi)^2}\frac{1}{2k^+}f^i_{lk}(k,q)f^i_{lm}(k,q)A^{a\dagger}_k(k^+,\v k)A^a_{m}(k^+,\v k)
O_1(q)=g^2N_c \int_{k:k^-<q^-} % \frac{dk^+}{2\pi} \frac{d^2  k}{(2\pi)^2}
\frac{1}{2k^+}f^i_{lk}(k,q)f^i_{lm}(k,q)
\langle \Psi_P | A^{a\dagger}_k(k^+, \v k)A^a_{m}(k^+,  \v k)|\Psi_P\rangle_E
\end{equation}
\begin{eqnarray}\label{2ps}
%O_2(q)&=&g^2\int_{\{(k-q)^-<q^-;k^-<q^-;\ (l-q)^-<q^-; l^-<q^-\}}\frac{dk^+}{2\pi} \frac{d^2\v k}{(2\pi)^2} \frac{dl^+}{2\pi} \frac{d^2\v l}{(2\pi)^2} f^i_{lk}(k,q)  f^i_{nm}(l,q)\nonumber\\
%&&:  A^{\dagger}_{l}(k^+,\v k-\v q)T^a  A_{k}(k^+,\v k) A^{\dagger}_{m}(l^+,\v l+\v q)T^a  A_{n}(l^+,\v l) :
O_2(q)
%&=&g^2\int_{\{(k-q)^-<q^-;k^-<q^-;\ (l-q)^-<q^-; l^-<q^-\}}\frac{dk^+}{2\pi} \frac{d^2  k}{(2\pi)^2} \frac{dl^+}{2\pi} \frac{d^2  l}{(2\pi)^2} f^i_{lk}(k-q,q)  f^i_{nm}(\v l,q)\nonumber\\ \nonumber \\
%&\times&\langle \Psi_P | :  A^{\dagger}_{l}(k^+,  k-  q)T^a  A_{k}(k^+,  k) A^{\dagger}_{m}(l^+,  l+  q)T^a  A_{n}(l^+,  l) :|\Psi_P\rangle\nonumber\\
&=&g^2\int_{\{k,l:\ (k-q)^-<q^-;k^-<q^-;\ (l-q)^-<q^-; l^-<q^-\}}%\frac{dk^+}{2\pi} \frac{d^2  k}{(2\pi)^2} \frac{dl^+}{2\pi} \frac{d^2  l}{(2\pi)^2}
 f^i_{lk}(k,q)  f^i_{nm}(l,q)\nonumber\\ \nonumber \\
&\times& \langle \Psi_P | :  A^{\dagger}_{l}(k^+,  \v k- \v q)T^a  A_{k}(k^+, \v k) A^{\dagger}_{m}(l^+, \v l+ \v q)T^a  A_{n}(l^+,\v  l) :|\Psi_P\rangle_E
\end{eqnarray}
Here $O_1$ represents a double scattering of a single projectile gluon, while $O_2$ corresponds to scattering of two distinct gluons from the projectile, each one scattering once.

Assuming rotationally invariant projectile wave function, we have
\begin{eqnarray}\label{o1}
O_1(q)&=&\frac{g^2N_c}{2} \int_{k^-<q^-}  \frac{dk^+}{2\pi} \frac{d^2\v k}{(2\pi)^2}\frac{1}{(2k^+)^2}f^i_{lk}(k,q)f^i_{lk}(k,q) \mathcal T(k)
%&=&\frac{g^2N_c}{2} \frac{1}{\v q^2}\int_{k^-<q^-}  \frac{dk^+}{2\pi} \frac{d^2\v k}{(2\pi)^2}\frac{1}{2k^+}\left[\frac{p^+(k^++p^+)}{k^+}\right]^24\frac{(k^+)^2+(k^++p^+)^2}{(p^+)^2}A^{a\dagger}_m(k^+,\v k)A^a_{m}(k^+,\v k)\nonumber
%&=&g^2N_c \frac{1}{\v p^2} \int_{k^-<p^-}  \frac{dk^+}{2\pi} \frac{d^2\v k}{(2\pi)^2}\frac{(k^++p^+)}{(k^+)^2}\left[(k^+)^2+(k^++p^+)^2\right]A^{a\dagger}_m(k^+,\v k)A^a_{m}(k^+,\v k)\nonumber\\
%&=&g^2N_c \frac{1}{\v p^2} \int_{k^-<p^-}  \frac{dk^+}{2\pi} \frac{d^2\v k}{(2\pi)^2}\frac{1}{2k^+}\frac{(k^++p^+)}{(k^+)^2}\left[(k^+)^2+(k^++p^+)^2\right]\hat T(k)
\end{eqnarray}
with
\begin{equation}\label{fsq}
[f^{Ti}(k,q)f^i(k,q)]_{st}=
\frac{(2k^+)^2}{\v q^2}\left[1+\frac{4}{(\frac{2q^-k^+}{\v q^2}-1)^2}\right]\delta_{st}
%=2\frac{1}{\v q^2}\left[\frac{k^++q^+}{k^+}\right]^2\left[(k^++q^+)^2+(k^+)^2\right]\delta_{st}
\end{equation}
and 
$\mathcal T$ - the gluon transverse momentum dependent distribution (TMD)
\begin{equation}\label{TMD}
%\hat T(k)\equiv  A^{\dagger ia}_i(k) A^a_i(k)
\hat T(k)\equiv  a^{\dagger a}_i(k) a^a_i(k);   \quad  \qquad \mathcal T(k)\equiv \langle \Psi_P|\hat T(k)|\Psi_P\rangle_E
\end{equation}
defined through the gluon number operator $\hat T$.

As noted in \cite{one}, this expression for the $S$-matrix reduces to the BFKL form for soft scattering, i.e. in the formal limit $\v q^2\ll 2q^-k^+$. In this limit the second term in \eqref{fijk} is negligible. Since $q^-$ is the frequency of the fastest gluon modes in the projectile, the combination $2q^-k^+$ is of the order of the transverse momentum of the projectile modes. Therefore the limit $\v q^2\ll 2q^-k^+$ indeed corresponds to soft scattering, where the momentum transfer from the target is smaller than the typical transverse momentum of gluons in the projectile wave function. 

On the other hand in the opposite limit $k^+\approx\frac{2\v q^2}{q^-}$ the second term dominates 
and  \eqref{fsq} can be approximated by\cite{one}
\begin{equation}\label{hscat}
 [f^{Ti}(k,q)f^i(k,q)]_{st}\approx
\frac{(2k^+)^2}{ \v q^2}\frac{1}{\epsilon}\delta(k^+-\bar q^+)\delta_{st}\,.
%=2\frac{1}{  q^2}\left[\frac{k^++q^+}{k^+}\right]^2\left[(k^++q^+)^2+(k^+)^2\right]\delta_{st}
\end{equation}
% \begin{equation}\label{hscat}
 %[f^{Ti}(k,q)f^i(k,q)]_{st}\approx
%\frac{(2k^+)^2}{\v q^2}\frac{4}{k^+}\delta[k^+-\frac{\v q^2}{2q^-}]\delta_{st}
%\end{equation}
where $\epsilon$ is related to the width of the target frequency distribution as $\epsilon=\Delta q^-/\v q^2$\cite{one}.
The single particle scattering part of the scattering amplitude (scaled by $\epsilon$) becomes
\begin{equation}\label{dis}
\bar O_1(q)=\frac{g^2N_c}{4\pi}\frac{1}{\v q^2} \int_{\v k^2<\v q^2=2k^+E}  \frac{d^2\v k}{(2\pi)^2}\,
\mathcal  T(\v k,k^+=\frac{\v q^2}{2q^-})
\end{equation}
%where we have identified $q^-=E$ as the frequency scale to which the projectile wave function has to be evolved.
Thus in the hard scattering limit
  \eqref{scat} together with \eqref{hscat} yields the DIS cross section (for a probe that interacts directly with gluons) at leading and subleading  twist. The leading twist part is given by $O_1$, while $O_2$ is  the subleading twist correction. 

In \cite{one} we have also shown that the BO evolved wave function upon fixed order perturbative expansion reproduces the large transverse logarithms encountered in the calculation of the NLO JIMWLK kernel in \cite{LuMu}.

In the present and following papers we turn our attention to explicit calculation of observables generated by the BO evolution. This paper is devoted to study of observables directly related to perturbative QCD calculations, most notably gluon TMD and gluon PDF. Our goal here is to discuss in detail the BO evolution of these quantities and to understand how it  relates to the standard perturbative CSS \cite{css1,css2,css3,css4} and DGLAP evolutions.

We note that gluon TMD's have been discussed in the framework of CGC before, see for example \cite{Marquet:2016cgx,Altinoluk:2018byz}. However as discussed in detail in \cite{jimwlkcss} the TMD's discussed in those work have longitudinal resolution scale equal to the longitudinal momentum of the gluon that is being measured by the TMD. They therefore are not subject to any evolution of the CSS type. Our interest in the present paper is more general, and we will study TMD at longitudinal resolution which is much different than the longitudinal momentum of the gluon (and similarly for the transverse resolution and the transverse momentum).

The paper is structured as follows. In Section 2 we discuss the  BO evolution of the gluon TMD. 
The gluon TMD defined in (\ref{TMD}) is the expectation value of the gluon number operator $\hat T(k)$
at fixed transverse momentum $\v k$ and fixed longitudinal momentum $k^+$ (or longitudinal momentum fraction $x\equiv k^+/P^+$).  The TMD is evolved in the logarithm of  frequency $\ln E/E_0$, which  plays the role of the resolution scale.
The BO evolution of the  TMD yields a nonlinear equation. We show that the linear term in this equation is closely related to the CSS equation. One has to keep in mind that the CSS evolution is a set of two equations with respect to the transverse and longitudinal resolution scales, $\mu^2$ and $\xi$ respectively. On the other hand,  the BO evolution has one evolution parameter only, $\ln E/E_0$.  We demonstrate that the (linearized) BO and CSS evolutions of the gluon TMD are equivalent if in the latter the longitudinal and transverse evolution parameters are not independent, but are related as $\ln \frac{\mu^2}{\v k^2}=2\ln \frac{k^+}{\xi}$
%where $\v q$ and $q^+$ are the longitudinal and transverse momenta of the gluon being resolved
\footnote{This relation holds in the doubly logarithmic approximation. To include single logarithms the relation is slightly modified - for details see Section 2.}. This aspect of our linearized calculation is closely related to \cite{BT1,BT2,BT3}, where the CSS equation also is recovered using the separation scale in $k^-$.

The nonlinear  in the TMD evolution equation in general cannot be expressed in terms of the TMDs and therefore with its inclusion the equation ceases to be closed. However in  a dilute approximation it does reduce to a term quadratic in the TMD. The physical meaning of the nonlinear term in this dilute approximation is quite straightforward to understand and is very interesting. It is a contribution of the stimulated emission: it enhances gluon emission through gluon splitting  into a bin in the phase space if the bin already contains gluons before the splitting occurs. 
As such the effect  is $1/N_c$ suppressed. At high transverse resolution scale $Q^2$, 
the stimulated emission effect is further suppressed by the factor $1/Q^2 S_\perp$,  where $S_\perp$ is the transverse area of the projectile.  This is the same suppression factor as in the so called GLR term \cite{GLR,MQ}, but the physics of the two is very different. In fact the stimulated emission correction is of the same order in $\alpha_s$ as the linear term, while the GLR term is suppressed by an additional power of $\alpha_s$. On the other hand the GLR term is enhanced al low $x$ by the logarithmic factor $\ln 1/x$, whereas the stimulated emission term does not exhibit such enhancement.

In Section 3 we analyze the BO evolution of the gluon PDF. Again we find that the equation has a linear and a nonlinear term.  In the leading logarithmic approximation, the linear term for not too small values of $x$ is identical to the DGLAP equation. It is interesting that the contribution to the real and virtual parts of the glue-glue splitting function have a little  different origin in the BO cascade as compared to the DGLAP/CSS cascade. This is due to different order of emission of some gluons in the two cascades.  In the DGLAP/CSS cascade gluons with highest transverse momentum and any longitudinal momentum are produced in the last step of the evolution. On the other hand the BO cascade is a little different:  at high enough longitudinal momentum the high transverse momentum gluons are produced by splittings in the middle of the cascade. Nevertheless, collecting all contributions we find that the resulting splitting functions in the BO approach coincide with the ones of the DGLAP. For very small $x$ we find an additional contribution to the linear part of the evolution. It originates from splittings which do not conform with the DGLAP kinematics (strong ordering in transverse momentum), but rather arise from the BFKL-like splittings, with strongly ordered longitudinal momenta. 
%These terms are simply encoded into the redefinition of the transverse resolution scale on the RHS of DGLAP equation from $Q^2$ to $Q^2/\zeta$ where $\zeta$ is the momentum fraction in the splitting.

The nonlinear contribution to the evolution again originates from the stimulated emissions. Interestingly, the leading %contribution from the stimulated emission enhancement turns out to be to 
nonlinear effect is in enhancement of  the virtual contribution to the DGLAP.  As a consequence the net nonlinear contribution to the evolution of the gluon PDF is negative and mimics shadowing corrections. We  stress again, that although the net effect has the same sign as the GLR shadowing correction, the physics here is very different and the parametric dependence on $\alpha_s$ and $\ln 1/x$ is different as well.

Finally we conclude in Section 4.
 
 The present paper is devoted to  study of the BO evolution of partonic observables. Those are observables that "live" in the middle of the cascade, i.e. envolve gluons that do not have the highest frequency available in the evolved wave function. The study of the observables of the JIMWLK, or BFKL type  - those that are dominated by the highest frequency gluons is left for the third paper  in this series \cite{three}. There we study the BO evolution of the total cross section for dilute-dilute scattering concentrating on eikonal emissions.  We derive and analyze the analog of the BFKL equation in this framework. We find that the frequency evolution has a very strong effect on the solutions of the BFKL equation slowing down the evolution of the scattering amplitude in a spectacular fashion.

\section{The evolution: the gluon TMD}
Our ultimate goal is to derive the evolution equation for the $S$-matrix. As a preamble to this, in the present paper we first consider  the evolution of the gluon TMD (\ref{TMD}). This is less cumbersome on one hand. On the other hand as we have observed in (\ref{S}), the TMD gives an important contribution to the $S$-matrix -- it is directly related to the term in the $S$-matrix associated with double scattering of a single projectile gluon, $O_1$. In addition it is an interesting observable in its own right,  and its evolution at high energy has recently been a focus of significant interest \cite{tmd1,tmd2,tmd3,tmd4,Caucal:2024bae}. 

The evolution of the gluon TMD within the JIMLWK formalism has been recently discussed in \cite{jimwlkcss}. 
The discussion was carried out within the standard $k^+$ evolution scheme employed in the standard JIMWLK evolution \cite{Bal1,Bal2,Bal3}, \cite{jimwlk1,jimwlk2,jimwlk3,jimwlk4,jimwlk5,jimwlk6},\cite{cgc1,cgc2,cgc3}.
We have found in \cite{jimwlkcss} that within the JIMWLK evolved state ($k^+$-ordered), for the TMD defined as the expectation value of the particle number operator $\hat T(k)$, the transverse resolution scale is automatically equal to the UV cutoff. This is  as long as the state is evolved in the longitudinal momentum past $k^+$ of the gluon in the TMD. This means that the JIMWLK TMD is not a very useful object, since in any physical observable the transverse resolution must be determined by the hard scale and not at all by the UV cutoff. If one wishes to express such a physical observable  in terms of the JIMWLK TMD one has to be prepared to perform additional resummation of large (ultraviolet) logarithms into the hard part of the process.

As we shall see in the BO approach the situation is very different.
Here the transverse resolution scale in the TMD is finite and related to the longitudinal resolution scale. This relation has precisely the same functional form that is needed to relate both resolutions scales to the  single hard scale in a physical way.  This means that in  the BO scheme  the hard part of a process must be free 
from large logarithms and would not require any additional resummation. 

% In addition we will find nonlinear corrections to the CSS evolution equation. These corrections superficially look  similar to the Gribov-Levin-Ryskin (or Mueller-Qiu) corrections to the evolution of PDF, but are distinct from these saturation corrections, and arise due to Bose nature of gluonic degrees of freedom.

 The TMD $\mathcal T(\v k^2,k^+)$ has been defined in (\ref{TMD}). 
%\beq\label{tmdd}
%{\mathcal T}(k^+,k^2;E)=\langle a^{\dagger\,a}_i(k) a_i^a(k)\rangle_E
%\eeq
The average of the gluon number operator is taken in the hadronic state \eqref{psiel1} evolved to frequency $E$. 
The frequency of the gluon is assumed to be smaller than the energy $k^-=\frac{\v k^2}{2k^+}<E$.
The evolution parameter $E$ provides the cutoff on the phase space of the gluons present in the wave function, and therefore provides the resolution scale for the TMD. One can view it as either the transverse resolution $\mu^2$ at fixed $k^+$ or the longitudinal resolution $\xi$ at fixed $\v k^2$. The exact relation of $E$ and the  resolution scales will be discussed later. 
We will also discuss in detail the relation between the evolution in the frequency $E$  and the CSS evolution equation.

% Defining $\mu^2$ as the transverse resolution scale, i.e. the maximal value of the transverse gluon momentum accessible in the wave function, and $\xi$ as the longitudinal resolution scale, i.e.  the minimal value  of the longitudinal momentum in the wave function, we have

%Note that in our definition the TMD has dimension $3$, as it is defined a particle density in three dimensional momentum space. To convert it to the standard definition which has dimension $2$ we need to multiply it by $P^+$ - the total longitudinal momentum of the hadron. This scaling does not affect the linear part of the evolution of TMD, but will be important in interpreting the nonlinear corrections, as we will see below.

%As we discuss below, the two resolution scales for the Born-Oppenheimer cascade are related by
%\beq\label{reso}
%\frac{\mu^2}{2\xi}=E%; \ \ \ \ \ \ \xi=\frac{k^2}{2E}
%\eeq
%Thus at fixed energy, the transverse and longitudinal resolutions in TMD defines in \eqref{tmdd} are determined by the longitudinal and transverse momenta respectively.

\subsection{The evolution - generalities}

%The evolution of the TMD is given entirely by the "Lindblad" or "virtual" part of the evolution. Fast gluons are emitted into the additional phase space opened by the evolution, and are averaged over since the observable itself does not involve these fast gluon degrees of freedom. The same contribution exists also for observables that do involve fast gluons, but for those there is an additional "real" contribution.

%This, as always arises from averaging over the highest frequency gluon emitted into the new "window". Note that in the "proper" DGLAP regime (i.e. NOT double log) both gluons arising after splitting of the gluon $k$, i.e. $p$ and $k-p$, are in the window. In this part of the evolution we will later average over the gluon $k-p$ in the vacuum. But for now we just keep the corresponding creation and annihilation operators in the game.

For any operator $\hat O$ consider the expectation value
\beq\label{expv}
\langle \hat O\rangle_E=\langle \Psi_P| \hat O| \Psi_P\rangle_E
\eeq
with $|\Psi_P\rangle_E$ given by \eqref{psiel1}. To derive the evolution equation of this expectation value, 
we introduce an   increment in the energy $Ee^\Delta$  for an infinitesimal $\Delta$. 
The interval of momenta 
 $E<p^-<Ee^\Delta$, which we will sometimes refer to  as "the window", 
 is the phase space opened by one small  step in the evolution.  The modes below $E$ are the valence modes while the window is populated by the fast modes, see Fig. \ref{Fig}.  The change in the expectation value due to a single step in the evolution is
 \beq\label{expv}
\delta \langle \hat  O\rangle\equiv \langle \hat O\rangle_{Ee^\Delta} -\langle \hat O\rangle_E
\eeq
%We write 
%\begin{equation}\label{psidelta}
%|\Psi_P\rangle_{Ee^{\Delta}}=e^{i\Delta EG(p^-=E)}|\Psi_P\rangle_{E}
%\end{equation}
%The energy increment  $\Delta E$ here is treated as infinitesimal. 
  The evolution is derived by  expanding the exponential factor in \eqref{psiel1} to second order in $C_i$ and averaging over the Hilbert space of the fast gluons. Since the state $|\Psi_P\rangle_{E}$ is a vacuum of the fast modes in the window,  
  $|\Psi_P\rangle_{E} =|0\rangle_F\otimes|\Psi_P\rangle_{E}$, we are able to explicitly average over these modes in calculating the expectation value.
 
\begin{figure}[H] % Use figure environment for captions
\centering % Center the plot
\begin{tikzpicture}
\begin{axis}[ scale=1.4,
        axis lines=left,             % Keep axes at the left and bottom
        axis line style={-latex},
	 xmin=0,   xmax=4,
	  ymin=0,   ymax=2.5,
         %extra x ticks={-1,1},
	 %extra y ticks={-2,2},
	 xtick=\empty,
	 xlabel={$k^-$}, % Optionally keep labels without numbers
           hide y axis, 
          x label style={at={(axis description cs:1,0)}}, 
        %  y label style={rotate=270, at={(axis description cs:0,1)}},
	%extra tick style={grid=major}, 
	clip=false,
	]    
       	%\addplot[dash pattern=on 5pt off 3pt] coordinates {(0,2) (4,2)};
	  \addplot[domain=0:0.8, dash pattern=on 5pt off 3pt] ({2},{x}); % A vertical line at x=2
	 \addplot[domain=0:0.8, dash pattern=on 5pt off 3pt] ({2.8},{x});
	 \node at (axis description cs:0.5,-0.1) [anchor=south] {$E$};
	 \node at (axis description cs:0.7,-0.1) [anchor=south] {$Ee^\Delta$};
	  \node[anchor=south east, font=\small] at (axis cs:1.6,0.4) {Virtual}; % Label for the cross
	  \addplot[only marks, mark=x, mark size=4pt, mark options={thick, black}] coordinates {(1.6, 0.4)};
	  
	  \node[anchor=south east, font=\small] at (axis cs:2.4,0.4) {Real}; % Label for the cross
	  \addplot[only marks, mark=x, mark size=4pt, mark options={thick, black}] coordinates {(2.4, 0.4)};
	 % \node at (axis description cs:-0.15,0.25) [anchor=west] {$X_{Bj}$}; 
	  % Here we add the brackets and labels
	   % \addplot[domain=0:2.8, smooth, thick, blue, mark=none] {2*(x/2.8)^2 * (2.8-x) * sin(pi*x/2.8)*500 / (1+x)^2} 
     %node[pos=0.4, above, font=\small, xshift=-15pt] {\textcolor{blue}{Projectile}};
    %---------------------------------------------------------------------------------------------------
       %\addplot[domain=2.8:4, smooth, thick, red, mark=none] {2 * (1 - (x - 2.8) / (4 - 2.8))^2 } 
       %     node[pos=0.2, left, yshift=5pt, font=\small, xshift=35pt] {\textcolor{red}{Target}};
% \addplot[domain=2.8:4, smooth, thick, red, mark=none] {(5)*exp(-20*(x-2.8))*(20*x-56)} 
   % node[pos=0.2, left, yshift=5pt, font=\small, xshift=50pt] {\textcolor{red}{Target}};    % \addplot[domain=0:2.8, smooth, thick, blue] {2*(x/2.8)^2 * (2.8-x) * sin(pi*x/2.8)*500 / (1+x)^2};
      {\color{red}
    \draw[decorate,decoration={brace,amplitude=5pt,mirror},very thick] (0,0) -- (2,0) node[midway,below=8pt,font=\small] {Slow/ Valence};
    \draw[decorate,decoration={brace,amplitude=5pt,mirror},very thick] (2,0) -- (2.8,0) node[midway,below=8pt,font=\small] {Fast};
    }
	\end{axis}
\end{tikzpicture}
\caption{A qualitative picture representing kinematics of virtual and real terms  for  the operator $\hat O$ averaged 
over the projectile wave function. The crosses indicate the momentum mode which builds the operator $\hat O$ in the "virtual" and "real" cases. }
\label{FigTMD} % Optional: label for referencing the figure
\end{figure}
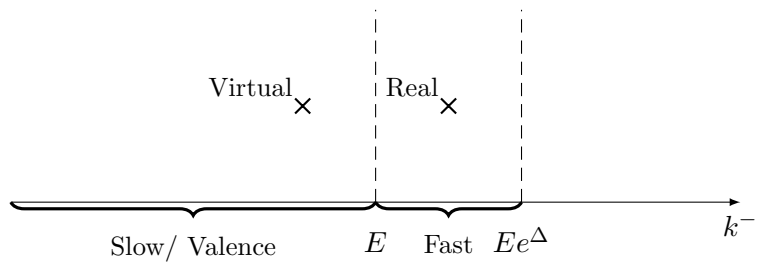

 Below we are going to consider two distinct cases. The first one is when the operator $\hat O$ by itself does not 
contain any gluon field modes inside the window. In this case (denoted as "virtual"  in Fig. \ref{FigTMD}),  the evolution is physically very similar to an evolution of an open quantum system in time, and the resulting equation is of the Lindblad type\cite{lindblad1,lindblad2}. The increment in the expectation value is naturally proportional to the total new phase space: 
$\delta \langle \hat  O\rangle \sim \Delta$. 
Hence, dividing by $\Delta$ we obtain a differential evolution:
 \beq
{ \partial \langle \hat  O\rangle \over \partial \eta} =\lim_{\Delta\rightarrow 0}{\delta \langle \hat  O\rangle\over \Delta}\,;
\,\qquad\qquad \eta\equiv \ln E/E_0.
 \eeq
 We will refer to this type of evolution as "virtual" since it is usually dominated by the decay of gluons in $\hat O$ by emitting gluons into the window\footnote{This is somewhat of a misnomer, as this evolution does contain real contribution as well, but it the examples considered in the present paper they are always smaller and only serve to cancel part of the phase spoce of virtual emissions.}.
  % with frequencies $E<k^-<Ee^\Delta$, for infinitesimally small $\Delta$. 
%Although the wave function is evolved with  the path ordered exponent in rapidity, this path ordering does not matter for our calculation. In our calculation we expand the evolution operator to second order only. In the first order, the path ordering clearly does not matter. In the second order term we only need the terms where the two frequencies are equal , and here again the path ordering does not matter. So we proceed as if there was no ordering in the evolution.

The second case is when $\hat O$  depends only on a gluon mode with momentu $p$ which is inside the window and $\langle \hat O\rangle_E=0$.
Hence $\delta \langle \hat  O\rangle= \langle \hat O\rangle_{Ee^\Delta}$. This average then is not proportional to $\Delta$, since  there is no integral over the phase space that could bring dependence on $\Delta$. We will refer to the evolution of such operators as "real" since the physical process that drives it, is the emission of the mode $p$ from slower modes below the window.

Generic operators of course depend on modes both below, and in the window but we will not deal with those in the present paper. We will meet such operators in \cite{three} and will adjust the calculation of their evolution accordingly.

\subsection{The "Lindblad (virtual) term": the gluon TMD at $k^-<E$.}

 We consider the gluon number operator \eqref{TMD}with $k$ such that  $k^-<E$.
To derive the evolution of any observable $\hat O$ of this type (which contains only creation and annihilation operators with frequencies $k^-<E$) the averaging  over the fast gluon Hilbert space in \eqref{expv} can be performed explicitly and in fact independently of the form of $\hat O$. This is because the only dependence on the fast glue creation and annihilation operators is contained in the wave function itself. The result is 
\begin{eqnarray}\label{lind}
\delta\langle \hat O\rangle&=&\int_{E<p^-<Ee^{\Delta}} \frac{d^3p}{(2\pi)^3}\frac{1}{2p^+}\langle C^\dagger (p)\hat OC(p)-\frac{1}{2}C^\dagger(p)C(p)\hat O-\frac{1}{2}\hat OC^\dagger(p)C(p)\rangle_E\nonumber\\
&=&\frac{1}{2}\int_{E<p^-<Ee^{\Delta}} \frac{d^3p}{(2\pi)^3}\frac{1}{2p^+}\langle\{C^\dagger (p)[\hat O,C(p)]+h.c.\}\rangle_E
\end{eqnarray}
%Here $\Delta$ is the (infinitesimal)  increment in the rapidity , and 
the factor $\frac{1}{2p^+}$ arises from the vacuum contraction (\ref{A}) of the high frequency fields. 
%$\langle 0| A(p)A^\dagger(p)|0 \rangle=\frac{(2\pi)^3}{2p^+}$. 
%The interval of momenta $E<p^-<Ee^\Delta$ is the phase space opened by one small (infinitesimal for $\Delta\rightarrow 0$) step in the evolution. We will sometimes refer to this momentum interval as "the window".

The above procedure is very reminiscent to averaging over unobserved part of the Hilbert space in the derivation of the evolution of reduced density matrix in open quantum systems. In that context the procedure leads (under certain favorable circumstances) to Lindblad evolution equation \cite{lindblad1,lindblad2}. By analogy we will call these terms in the evolution the "Lindblad terms". 
If the operator $\hat O$ does contain fast gluon creation/annihilation operators, the terms \eqref{lind} still arise, although the evolution equation will also contain additional, "real"  terms. Such terms arise for example  when  considering the TMD at the very edge of the available phase space, i.e. for gluons with  $k^-\simeq E$. We will explore this case separately below.

Note that if it were true that $C^\dagger(p^+,\v p)=C(p^+,-\v p)$, we could reorganize the right hand side of (\ref{lind}) into a double commutator. This is precisely what happens in the standard calculation with $k^+$ ordering,  where one neglects the dependence on  $p^+$. This leads to significant simplification, since the double commutator preserves the number of creation and annihilation operators in $\hat O$. Thus in the $k^+$-ordered scheme, the Lindblad evolution of any operator $\hat O$ yields a homogeneous evolution equation. In particular, in \cite{jimwlkcss}  the equation for the TMD (and similar operators) was derived  resulting in the CSS evolution equation  with respect to the longitudinal cutoff with the transverse resolution scale being equal to the UV cutoff.

%For more general operators, that possibly depend on the fast degrees of freedom this statement remains true. If we were able to represent the RHS of \eqref{lind} as a double commutator, it would ensure that the Lindblad part of the evolution  was homogeneous, i.e. that the number of creation and annihilation operators  on the RHS of \eqref{lind} is the same as in the operator $O$ itself. In particular the Lindblad part of the evolution of  the single gluon scattering amplitude $O_1$ of \eqref{o1}  would be homogeneous, as would also be that of the two gluon scattering  amplitude $O_2$ separately.

However in the full BO calculation one cannot neglect the dependence of $C$  on $p^+$. The relation $C^\dagger(p^+,\v p)=C(p^+,-\v p)$ then cannot be true by momentum conservation, since $C^\dagger(p^+)$ creates momentum $p^+$, while $C(p^+)$ annihilates momentum $p^+$. 
It is strongly violated by the DGLAP contribution where $p^+$ cannot be neglected in any approximation. As a result even in the dilute  projectile limit, which we are considering in this paper, the evolution equations we obtain are not homogeneous and contain nonlinear terms. We will organize the result of the calculation by normal ordering the right hand side of the evolution equation to separate the homogeneous and the inhomogeneous terms.

Finally we note that in the following we first calculate the change in observable due to the finite width of the window, and then take the limit $\Delta\rightarrow 0$ to turn this into a differential evolution equation in rapidity. In order to go from the calculation of the finite increment to differential equation we
represent the integral over $p$ within the window in the limit of $\Delta\rightarrow 0$ as
\begin{eqnarray}\label{intwin}
\int\frac{dp^+d^2\v p}{(2\pi)^3}f(p)&=&\int \frac{dp^+}{2\pi}\frac{1}{4\pi}\int dp^2f(p)=\Delta\frac{1}{4\pi}\int \frac{dp^+}{2\pi}2p^+E f(\v p^2=2p^+E,p^+)\\
&=&\Delta\frac{1}{2\pi}\int\frac{d^2\v p}{(2\pi)^2}\frac{\v p^2}{2E}f(\v p^2, p^+=\frac{\v p^2}{2E})\nonumber
\end{eqnarray}
where we have used the fact that for $\Delta\rightarrow0$ we can set $p^-=E$. The differential from of the evolution equation for the expectation value in \eqref{expv} is then obtained  simply by dropping the factor $\Delta$ in \eqref{intwin}.

A direct calculation for the evolution of the gluon TMD gives the following result
\begin{eqnarray}\label{evv}
\delta\mathcal{T}(k)\equiv  \langle \hat T\rangle_{Ee^\Delta} -\langle \hat T\rangle_E\,;\qquad\qquad
\delta\mathcal{T}(k)=\delta\mathcal{T}^L(k)+\delta\mathcal{T}^{NL}(k)
\end{eqnarray}
with
\begin{eqnarray}\label{tmd1e}
\delta\mathcal{T}^L(k)=\frac{g^2N_c}{2}\int_{k^-, (k\pm p)^-<E;E<p^-<Ee^{\Delta}}  \int \frac{d^3p}{(2\pi)^3}\frac{1}{2p^+}\Big[&&\frac{1}{4k^+(k^++p^+)}F^{l}_{st}(k+p,p)F^l_{st}(k+p,p)\mathcal{T}(k+p)\nonumber\\
&&-\frac{1}{4k^+(k^+-p^+)}F^l_{st}(k,p)F^l_{st}(k,p)\mathcal{T}(k) \Big]
\end{eqnarray}
and 
\begin{eqnarray}\label{tmd2}
&&\delta \mathcal{T}^{NL}(k)=\frac{g^2}{2}\int\frac{d^3p}{(2\pi)^3}\int\frac{d^3l}{(2\pi)^3}\frac{1}{2p^+}\\
&&\times \Bigg[
F^i_{lk}(l,p)F^i_{nm}(k+p,p) \langle\Psi_P|: A^{\dagger}_{l}(l^+-p^+,\v l-\v p)T^a  A_{k}(l^+,\v l)
 A^{\dagger}_{m}(k^++p^+,\v k+\v p)T^a  A_{n}(k^+,\v k):|\Psi_P\rangle_E\nonumber\\
&&-F^i_{lk}(l,p)F^i_{nm}(k,p) \langle\Psi_P|: A^{\dagger}_{l}(l^+-p^+,\v l-\v p)T^a  A_{k}(l^+,\v l)
 A^{\dagger}_{m}(k^+,\v k)T^a  A_{n}(k^+-p^+,\v k-\v p):|\Psi_P\rangle_E\Bigg]+h.c.\nonumber
\end{eqnarray}
where the superscripts $L$ and $NL$ denote the linear and nonlinear terms.
In arriving at this form we have normal ordered the term that contains a product of four creation and annihilation operators. Physically this is convenient since such normal ordering separates the one particle operators from two particle operators in a natural way. We note that in the eikonal approximation, i.e. assuming $p^+\ll l^+$ the nonlinear term vanishes, as it should in the BFKL limit. 

We will now discuss these expressions in more detail.
We start with (\ref{tmd1e}) -- the linear term 
in the evolution equation, and will return to the physical interpretation of the nonlinear corrections (\ref{tmd2}) later.

\subsubsection{The linear term in the evolution}

The change of the TMD is due to opening of the new phase space between $E$ and $Ee^\Delta$. 
The homogeneous piece  is given by \eqref{tmd1e}. 
%\begin{eqnarray}
%\delta\langle O^L_1(q)\rangle&=&g^4N_c^2\int_{k^-<p^-; \ (k-p)^-<p^-} \frac{d^3k}{(2\pi)^3}\langle A^{a\dagger}_i(k)A^a_j(k)\rangle\int \frac{d^3p}{(2\pi)^3}\frac{1}{2p^+}\left[F^{lT}(k,p)F^l(k,p)\right]_{ij}\\
%&\times&\left[\frac{1}{8(k^+-p^+)^2(k^+-p^++q^+)}f^m_{nk}(k-p,q)f^m_{nk}(k-p,q)-\frac{1}{8k^+(k^+-p^+)(k^++q^+)}f^m_{kn}(k,q)f^m_{kn}(k,q)\right]\nonumber
%\end{eqnarray}
%This can be written in a way that factorizes the $q$ and $p$ dependence by changing variables $k\rightarrow k+p$ in the first term. Here we also assume rotational invariance of the wave function
%\begin{eqnarray}\label{tmd1e}
%\delta\mathcal{T}^L(k)=\frac{g^2N_c}{2}\int_{k^-, (k\pm p)^-<E;E<p^-<Ee^{\Delta}}  \int \frac{d^3p}{(2\pi)^3}\frac{1}{2p^+}\Big[&&\frac{1}{4k^+(k^++p^+)}F^{l}_{st}(k+p,p)F^l_{st}(k+p,p)\mathcal{T}(k+p)\nonumber\\
%&&-\frac{1}{4k^+(k^+-p^+)}F^l_{st}(k,p)F^l_{st}(k,p)\mathcal{T}(k) \Big]
%\end{eqnarray}
%In the limits of the integration in \eqref{o1e} the $+$ sign refers to the first term, while the $-$ sign - to the second term.
The first term here is the gain term originating from splitting of gluon with momentum $k+p$ into a gluon in the window and a gluon with momentum $k$, while  the second  is a loss term where the gluon with momentum $k$ decays and thus disappears from the TMD. 
The loss term is obviously the virtual term in usual terminology, while the gain term is % (one  contribution to) 
 a real term.
To bring this equation into a familiar form we use the explicit form of $F$ \eqref{F} to find
\begin{equation}
F^i_{ln}(k,p)F^i_{ln}(k,p)=32 (k^+)^2\zeta(1-\zeta)\left[\frac{\zeta}{1-\zeta}+\frac{1-\zeta}{\zeta}+\zeta(1-\zeta)\right]\frac{1}{\v p^2}; \ \ \ \ \ \ \ \zeta\equiv\frac{p^+}{k^+}
\end{equation}
When integrating the second (loss) term in \eqref{tmd1e} we need to impose the constraint $(k-p)^-<E$ since in our nomenclature  the gluon with momentum $p$ always has the highest frequency. For the BFKL regime, $p^+\ll k^+$ this is satisfied once $k^-<E$. For the DGLAP part of phase space, $\v p^2\gg \v k^2$, this constraint is equivalent to $\zeta<1/2$. Thus the integral over $\zeta$ is bounded by $1/2$. 
We then find
\begin{equation}\label{loss1}
\int \frac{d^3p}{(2\pi)^3}\frac{1}{2p^+}\frac{1}{4k^+(k^+-p^+)}F^l_{st}(k,p)F^l_{st}(k,p)\mathcal{T}(\v k,x)=\frac{\Delta}{2\pi^2}\int_0^{1/2}d\zeta\left[\frac{\zeta}{1-\zeta}+\frac{1-\zeta}{\zeta}+\zeta(1-\zeta)\right]\mathcal{T}(\v k,x)
\end{equation}
where 
%\begin{equation}
$x\equiv k^+/P^+$ which we introduce as an argument of $\mathcal{T}$
 instead of the longitudinal momentum $k^+$\footnote{One can easily check that $\mathcal{T}(\v k, k^+)\equiv P^+\mathcal{T}(\v k,x)$}.
%\end{equation}
%with $P^+$ - the total longitudinal momentum of the projectile. 

For the gain term we get 
\begin{eqnarray}\label{gain1}
&&\int \frac{d^3p}{(2\pi)^3}\frac{1}{2p^+}\frac{1}{4k^+(k^++p^+)}F^l_{st}(k+p,p)F^l_{st}(k+p,p)\mathcal{T}(\v k+\v p,k^++p^+)
\\
&&
\qquad\qquad =\frac{\Delta}{2\pi^2}\int d\v n\int_0^{1/2}d\zeta\left[\frac{\zeta}{1-\zeta}+\frac{1-\zeta}{\zeta}+\zeta(1-\zeta)\right]\frac{1}{1-\zeta}\mathcal{T}(\v k_\zeta,\frac{x}{1-\zeta})\nonumber
\end{eqnarray}
where now $\zeta\equiv\frac{p^+}{k^++p^+}$ and
\begin{equation}
\v k_\zeta\equiv \v k+\v  n\left[2Ek^+\frac{\zeta}{1-\zeta}\right]^{1/2}
\end{equation}
where $\v n$ is a unit vector and the integral is over the angle of $\v n$ in the transverse plane (normalized to unity).

In total we have
\begin{eqnarray}
\frac{\partial}{\partial \eta}\mathcal{T}(\v k, x)&=&-\frac{g^2N_c}{4\pi^2}\int_{0}^{1/2}d\zeta\left[\frac{\zeta}{1-\zeta}+\frac{1-\zeta}{\zeta}+\zeta(1-\zeta)\right] \left[\mathcal{T}(\v k, x)-\frac{1}{1-\zeta}\int d\v n\;\mathcal{T}(\v k_\zeta,\frac{x}{1-\zeta})\right]
 \end{eqnarray}
The gain term of \eqref{gain1} regulates the $\zeta\rightarrow 0$ divergence in the loss term \eqref{loss1}. As long as $\v k^2\gg 2k^+E\frac{\zeta}{1-\zeta}$, or
\begin{equation}\label{cutoff1}
\zeta<\frac{k^-/E}{1+k^-/E}\approx \frac{k^-}{E}
\end{equation}
 there is a cancellation between the two terms.
When $k^-\ll E$ the cancellation only provides the rapidity regulator, while for $k^-\sim E$ the cancellation is almost complete. For larger values of $\zeta$ the argument of the TMD in \eqref{gain1} is large. Making a natural assumption that the TMD is a decreasing function of transverse momentum, we can neglect the gain term for these values of $\zeta$. This means that the only purpose of the gain term is to provide a cutoff for the loss term at the value of $\zeta$ given by \eqref{cutoff1}. 

Thus we can write
\begin{eqnarray}\label{linear}
\frac{\partial}{\partial \eta}\mathcal{T}(\v k, x)&=&-\frac{g^2N_c}{4\pi^2}\int_{\frac{k^-}{E}}^{1/2}d\zeta\left[\frac{\zeta}{1-\zeta}+\frac{1-\zeta}{\zeta}+\zeta(1-\zeta)\right] \mathcal{T}(\v k, x)+...
 \end{eqnarray}
% {\color{red} $\frac{g^2N_c}{2\pi^2} \rightarrow \frac{g^2N_c}{4\pi^2}$. An extra factor of $1/2$ is from the $\frac{g^2N_c}{2}$ in Eq.~\eqref{tmd1e}}
where ellipsis refers to the nonlinear term which we will deal with in a little while.

Interestingly, the emissions that cancel between the loss and the gain terms correspond to splittings $\v k+\v p\rightarrow (\v k, \v p)$ and $\v k\rightarrow (\v k-\v p, \v p) $
with $\v p^2<\v k^2$, i.e. to splittings that do not conform to the DGLAP kinematics of strong transverse momentum ordering. Thus the actual evolution of the TMD \eqref{linear} is driven by the DGLAP splittings alone.

\subsubsection{The CSS equation and the issue of resolution scales}

The form of the evolution equation \eqref{linear} is very close to the CSS equation, and so is its physical meaning.  Recall that we are evolving in frequency past the frequency $k^-$ at which we are counting the number of gluons with the number operator $\hat T$. This evolution creates additional gluons in the wave function with transverse momenta $\v p^2>\v k^2$ and longitudinal momenta $p^+<k^+$, which arise from perturbative splittings of the gluon at $k$. This is precisely the same mechanism of increasing transverse and longitudinal resolution of a TMD that leads to  the CSS evolution equation, which operates strictly within the DGLAP/CSS cascade (strong transverse momentum ordering). Thus we expect \eqref{linear}  to be related to the CSS equation. We now discuss the nature of this relation.

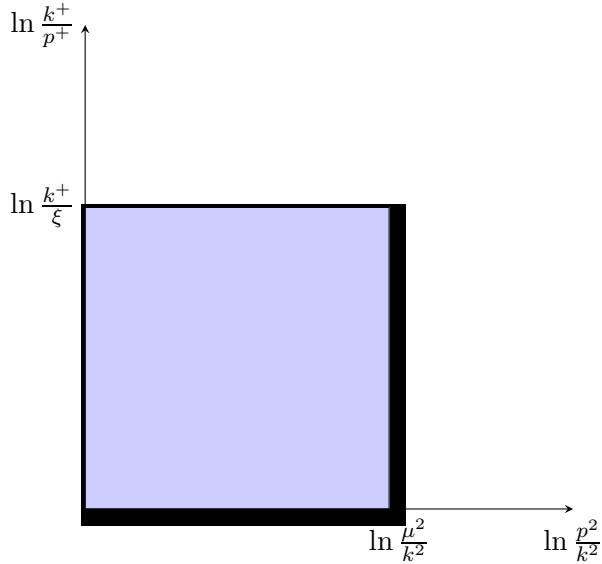
\begin{figure}
\centering
\begin{tikzpicture}
         % Draw a square with internal shadow
    \filldraw[draw=black, fill=blue!20, blur shadow={shadow blur steps=5}] (0,0) rectangle (4,4);
     % \node at (0,0) [anchor=north east] {A}; % Lower-left corner
     \node at (3.6,0) [anchor=north west] {$\ln \frac{\mu^2}{k^2}$}; % Lower-right corner
     %\node at (6,6) [anchor=south west] {C}; % Upper-right corner
     \node at (0,3.6) [anchor=south east] {$\ln \frac{k^+}{\xi}$}; % Upper-left corner
    % Optionally add the axes and labels
    \begin{axis}[
        axis x line=bottom,
        axis y line=left,
        xtick=\empty,
	ytick=\empty,
       xlabel={$\ln \frac{p^2}{k^2}$}, 
        ylabel={$\ln \frac{k^+}{p^+}$},
        xmin=0, xmax=4,
        ymin=0, ymax=4,
        width=8cm, height=8cm, % Ensures the square remains square-shaped
        axis equal,
        clip=false,
            x label style={at={(axis description cs:1,0)}}, 
          y label style={rotate=270, at={(axis description cs:0,1)}},
    ]
    \end{axis}
     \end{tikzpicture}
     \caption{The phase space of the DGLAP/CSS cascade with resolution scales $\mu^2$ and $\xi$. Only gluons with momenta $(p^+,\v p^2)$ inside the blue rectangle are allowed in the wave function.}
\label{f1}
\end{figure}

To recap,  the basic physics of the CSS evolution is very simple. Consider a {\bf single} parton with momentum $(\v k,k^+)$. At this point (the initial point of the evolution) the transverse and the longitudinal resolution with which we are resolving  the parton are obviously $\mu_0^2=\v k^2$ and $\xi_0=k^+$\footnote{Here we use $\xi$ of dimension one, as opposed to the standard CSS notations where it is of dimension two, and corresponds to $(k^+)^2$.}, since the parton is not dressed by any emissions. Let us now include DGLAP (strongly transverse momentum ordered) splittings in the wave function of the initial parton, but limit the available phase space in the transverse momentum-longitudinal momentum plane by a rectangle with sides $\mu^2$ and $\xi$ (see Fig. \ref{f1}). That is to say we  cutoff the part of the cascade  at momenta of the emitted gluons $p$ by $\v p^2<\mu^2$, $p^+<\xi$. This we  refer to as the DGLAP/CSS cascade. On a logarithmic plot,
the DGLAP/CSS cascade populates the rectangle in the $\ln (\frac{\v k^2}{\v p^2})-\ln (\frac{p^+}{k^+})$ plane, see Fig. \ref{f1}. Varying these scales, i.e. allowing more or less DGLAP emissions  leads to the following 
evolution equations (CSS), provided $\mu^2\gg\mu_0^2$ and $\xi\ll \xi_0$ 
\begin{equation}
\begin{split}\label{css}
\frac{\partial  T(k^+,\v k^2;\mu^2; \xi)}{\partial \ln \mu^2}=&-
\frac{\alpha_s}{2\pi}N_c\int_{\xi/k^+}^{1-\xi/k^+}d\zeta \Big[\frac{1-\zeta}{\zeta}+\frac{\zeta}{1-\zeta}+\zeta (1-\zeta) \Big]\,T\left(k^+,\v k^2;\mu^2; \xi\right) \\
\frac{\partial  T(k^+,\v k^2;\mu^2; \xi)}{\partial \ln \frac{1}{\xi}}=& -
\frac{\alpha_s}{2\pi}2N_c \int_{\v k^2}^{\mu^2}\frac{dp^2}{p^2}T\left(k^+,\v k^2;\mu^2; \xi\right)
\end{split}
\end{equation}
Here we have introduced $T(k^+,\v k^2;\mu^2; \xi)$ as a TMD evolved by the CSS, to distinguish it form  $\mathcal{T}$ defined 
earlier.

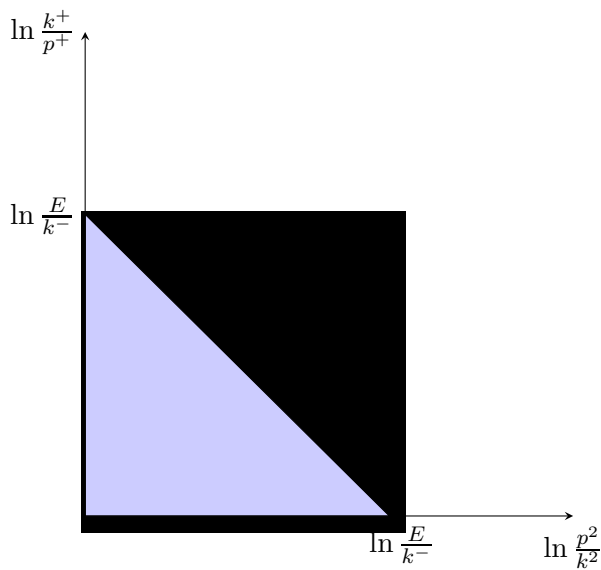
\begin{figure}[t]
\centering
\begin{tikzpicture}
         % Draw a square with internal shadow
    \filldraw[draw=black, fill=blue!20, blur shadow={shadow blur steps=5}] (0,0) -- (4,0) -- (0,4) -- cycle;     % \node at (0,0) [anchor=north east] {A}; % Lower-left corner
     \node at (3.6,0) [anchor=north west] {$\ln \frac{E}{k^-}$}; % Lower-right corner
     %\node at (6,6) [anchor=south west] {C}; % Upper-right corner
     \node at (0,3.6) [anchor=south east] {$\ln \frac{E}{k^-}$}; % Upper-left corner
    % Optionally add the axes and labels
    \begin{axis}[
        axis x line=bottom,
        axis y line=left,
        xtick=\empty,
	ytick=\empty,
       xlabel={$\ln \frac{p^2}{k^2}$}, 
        ylabel={$\ln \frac{k^+}{p^+}$},
        xmin=0, xmax=4,
        ymin=0, ymax=4,
        width=8cm, height=8cm, % Ensures the square remains square-shaped
        axis equal,
        clip=false,
            x label style={at={(axis description cs:1,0)}}, 
          y label style={rotate=270, at={(axis description cs:0,1)}},
    ]
    \end{axis}
     \end{tikzpicture}
     \caption{The phase space of the BO cascade. Only gluons with $p^-<E$, $p^+<k^+$ and $\v p^2>\v k^2$ are present in the wave function. }
     \label{f2}
\end{figure}

To understand the relation between the BO approach and the CSS evolution, we first note the obvious: the CSS has two resolution parameters and therefore two distinct equations, whereas we have introduced only one "regulator" - the total frequency E in the BO. Thus in the BO formalism,  clearly we can only ever have one evolution equation. On the other hand at finite $E$ the evolved wave function does not contain gluons with arbitrary longitudinal and transverse momenta, but only those satisfying the condition $\frac{\v p^2}{2p^+}<E$. 
In other words we can expect that the TMD $\mathcal{T}$ calculated as the particle number in BO evolved wave function  is related to the CSS evolved TMD $T$ as
\beq\label{corresp}
\mathcal{T}(\v k,k^+;E)=T(k^+,\v k;\mu^2(E);\xi(E))
\eeq
for some functional dependence $\mu^2(E),\ \xi(E)$. To determine the functional dependence $\mu^2(E), \ \xi(E)$ we have to understand the phase space of the BO cascade. This is depicted on Fig. \ref{f2}. The phase space populated by gluons is defined by conditions $\v p^2/2p^+<E$, $p^+>k^+$, $\v p^2>\v k^2$. Thus (in logarithmic coordinates)  the BO cascade populates  triangle with sides of length $\ln \frac{E}{k^-}$.
%\begin{figure}[t]         
%\centering                              
   %                               \includegraphics[width=6cm] {CSS vs BO.pdf}             
%\caption{The respective phase spaces of the CSS and BO evolutions}
%\label{f1}
%\end{figure}
%\begin{figure}[t]         
%\centering                              
   %                               \includegraphics[width=6cm]{CSS.png}             
%\caption{The phase space of the CSS cascade}
%\label{f1}
%\end{figure}

%\begin{figure}[t]         
%\centering                              
   %                               \includegraphics[width=6cm]{CSS_BO.png}             
%\caption{The phase space of the BO cascade.} %Comparison between BO evolutions and CSS evolution that satisfy $\frac{\mu^2}{\xi}=2E$ and maximize the overlap with BO.}
%\label{f2}
%\end{figure}
The phase space of the BO cascade is therefore different from the DGLAP/CSS cascade. An additional difference between the two is that in the BO approximation the cascade in principle also contains BFKL type emissions, in addition to the DGLAP splittings. However as we have seen in the previous subsection, the splittings with $\v p^2<\v k^2$ cancel between the gain and loss terms in the BO TMD evolution, while for $\v p^2>\v k^2$ the BFKL contribution is limited to the doubly logarithmic kinematics, which is also part of the DGLAP contribution, and so in this respect the difference is of no consequence.

In spite of the different phase space, one can identify a DGLAP/CSS cascade which is equivalent to the BO cascade as far as the evolution of TMD is concerned. First let us consider the doubly logarithmic approximation.  In the doubly logarithmic approximation the DGLAP emission probability has a simple property: the probability of emission into a unit area of phase space (in logarithmic coordinates) does not depend on the location in the phase space. This immediately implies that the total probability of emission does not depend on the shape, but only on the total area of the phase space populated by the cascade. Thus choosing the following parameters for the DGLAP/CSS cascade
\beq\label{reso}
\ln \frac{\mu^2(E)}{\v k^2}=\frac{1}{\sqrt 2}a\ln \frac{E}{k^-}; \ \ \ \ \ \ln \frac{k^+}{\xi(E)}=\frac{1}{\sqrt 2}\frac{1}{a} \ln \frac{E}{k^-}
\eeq
with an arbitrary constant $a$, the DGLAP/CSS cascade and the BO cascade yield identical TMD's, in the doubly logarithmic approximation.
%To extend this equivalence beyond the doubly logarithmic approximation one can use the freedom in the choice of $a$, but we will not pursue this in this paper.

Indeed using \eqref{reso} in \eqref{corresp}, and keeping only leading logarithmic contributions (double logarithms in the standard Sudakov terminology)  in the integral in \eqref{css} we find that \eqref{linear} indeed yields
\beq\label{corr}
\frac{\partial}{\partial \ln E}\mathcal{T}(\v k,k^+;E)=\left[\frac{\partial \ln \mu^2}{\partial \ln E}\frac{\partial}{\partial \ln \mu^2}+\frac{\partial \ln 1/\xi}{\partial \ln E}\frac{\partial}{\partial \ln 1/\xi}\right]T(\v k,k^+;\mu^2(E),\xi(E))
\eeq
We note that $a=\sqrt{2}$ is the natural choice for the constant in \eqref{reso}. It leads to the relation between the transverse and longitudinal factorization scales $\ln \frac{\mu^2}{\v k^2}=2\ln \frac{k^+}{\xi}$.This relation is necessary in the standard CSS approach to ensure that the physical Sudakov suppression form factor in  processes with one hard scale $Q^2$ is fully attributable to the evolution of the TMD, with no additional large logarithms arising in the hard part of the cross section. We will also see below that this choice arises naturally when considering the gluon PDF within the BO cascade.

 To include the single logarithmic term we have to keep a constant ($\xi$-independent) term in the integrand on the RHS of the first of the CSS equations \eqref{css}. This modifies the picture slightly. Now the emission probability is flat in the coordinates $\ln \frac{k^+}{p^+}-\frac{11}{12}$ and $\ln \frac{\v p^2}{\v k^2}$. The condition of the equivalence of the two
cascades including the single logarithmic terms is
\beq\label{resol}
\ln \frac{\mu^2(E)}{\v k^2}=\frac{1}{\sqrt 2}a\ln \frac{E}{k^-}; \ \ \ \ \ \ln \frac{k^+}{\xi(E)}-\frac{11}{12}=\frac{1}{\sqrt 2}\frac{1}{a} \left[\ln \frac{E}{k^-}-\frac{11}{12}\right]
\eeq
One can check explicitly that now \eqref{corr} holds including the single logarithmic terms.

We have thus established that the linear part of the BO evolution  for the gluon TMD is completely equivalent to the CSS equation where the transverse and longitudinal resolution scales are chosen to depend on energy according to \eqref{resol}.  We note that this is very different from the TMD evolution in the standard JIMWLK approach where the evolution parameter is the longitudinal momentum rather than frequency. As we showed in \cite{jimwlkcss}, the evolution of the gluon TMD in the JIMWLK framework is indeed equivalent to the (doubly logarithmic) CSS equation, but with the transverse resolution scale 
being fixed at the UV cutoff. This is because in $k^+$ evolution the transverse and longitudinal coordinates are completely decoupled and emission of arbitrarily high transverse momentum gluons is allowed in one step of the evolution. Although there is fundamentally nothing wrong with such evolution, it renders the notion of the evolved gluon TMD practically useless for applications:  the UV divergence present in this evolution has to be cancelled in physical observables by a similar (divergent) contribution to the appropriate "hard part" of the process. In this respect the BO evolution is much more physical since, as expected, it is equivalent to the CSS evolution with finite transverse and longitudinal resolution scales, which eventually are both set by the hard scale in the observable that is being measured.

This concludes our discussion of the correspondence between the linear part of the BO and CSS evolutions of the gluon TMD.
% In view of this correspondence, in the following we will refer to \eqref{linear} as the CSS-BO equation.
We now turn to the discussion of nonlinear terms in the evolution equation \eqref{evv}.

\subsubsection{ The nonlinear contribution -- the stimulated emission correction}

In addition to the homogeneous term (\ref{tmd1e}), the evolution equation contains $\delta\mathcal{T}^{NL}$ --
a nonlinear contribution (\ref{tmd2}). Below we discuss the physics behind this term.
%\begin{eqnarray}\label{nonlin}
%&&\delta \mathcal{T}^{NL}(k)=\frac{g^2}{2}\int\frac{d^3p}{(2\pi)^3}\int\frac{d^3l}{(2\pi)^3}\frac{1}{2p^+}\\
%&& \Bigg[
%F^i_{lk}(l,p)F^i_{nm}(k+p,p) \langle\Psi_P|: A^{\dagger}_{l}(l^+-p^+,\v l-\v p)T^a  A_{k}(l^+,\v l)
 %A^{\dagger}_{m}(k^++p^+,\v k+\v p)T^a  A_{n}(k^+,\v k):|\Psi_P\rangle\nonumber\\
%&&-F^i_{lk}(l,p)F^i_{nm}(k,p) \langle\Psi_P|: A^{\dagger}_{l}(l^+-p^+,\v l-\v p)T^a  A_{k}(l^+,\v l)
 %A^{\dagger}_{m}(k^+,\v k)T^a  A_{n}(k^+-p^+,\v k-\v p):|\Psi_P\rangle\Bigg]+h.c.\nonumber
%\end{eqnarray}
%Again, we note that in the eikonal approximation, i.e. assuming $p^+\ll l^+$ this nonlinear term vanishes, as it should in the BFKL limit.

\subsubsection*{The dilute approximation}
Written in the form (\ref{tmd2})  the nonlinear term is not particularly illuminating. We will encounter other similar nonlinear terms later in our analysis which also are quite obtuse. One can however understand the basic physics of these terms by considering the following simple approximation, which we will call the "dilute approximation".

 Suppose the wave function of the projectile is dilute, so that at every value of momentum the state contains at most one particle.
 Let us also 
 %first 
 assume, for the sake of the argument, that the hadronic wave function $|\Psi_P\rangle_E$ contains a fixed number of particles (zero or one) at any given value of momentum, in other words has a form $ |k_1, k_2,...k_n\rangle$, and is of course a color singlet.   In this case the average of four gluon field operators simplifies to
\begin{eqnarray}\label{diluteaverage1}
&&\langle \Psi_P|: A^{\dagger b}_{l}(k^+-p^+,\v k-\v p)T^a_{bc}  A^c_{k}(k^+,\v k)
 A^{\dagger d}_{m}(l^++p^+,\v l+\v p)T^a_{de}  A^e_{n}(l^+,\v l):|\Psi_P\rangle_E\\
&&=\frac{1}{4V(N_c^2-1)^2}\delta^3(k-l-p)\delta^{km}\delta^{ln}\delta^{cd}\delta^{be}T^a_{bc}T^a_{de} \nonumber \\
&&\times\langle\Psi_P|:\left[ A^{\dagger }(l^+,\v l) A(l^+,\v l)\right]\left[ A^\dagger(l^++p^+,\v l+\v p)
 A(l^++p^+,\v l+\v p) \right]:|\Psi_P\rangle_E\nonumber\\
&&=\frac{N_c}{4V(N_c^2-1)}\delta^3(k-l-p)\delta^{km}\delta^{ln}\langle\Psi_P|:\left[ A^{\dagger }(l^+,\v l) A(l^+,\v l)\right]\left[ A^\dagger(l^++p^+,\v l+\v p)
 A(l^++p^+,\v l+\v p) \right]:|\Psi_P\rangle_E\nonumber
\end{eqnarray}
where the square brackets denote scalar product of the gluon fields in color and polarization.
This equation is simply the statement that the state of the incoming two gluons must be exactly the same as that of the outgoing two gluons (with the assumption of rotational and color invariance of the state), otherwise the expectation value vanishes. The factor $V$ here is the three dimensional volume which appears because the expectation value is inversely proportional to $\delta^3(p=0)\equiv V$. 

The state above ($|\Psi_P\rangle= |k_1, k_2,...k_n\rangle$) is however far from generic. A generic state would contain a superposition of gluons with different values of momenta. Thus in general the two annihilated gluons do not have to have exactly the same momenta as the two created gluons. 
Nevertheless \eqref{diluteaverage1} is a useful approximation for a certain type of wave functions/momenta.

To understand the basics let us consider for simplicity  a two particle state  
\begin{equation}
|\Phi\rangle=\int_{\v q,\v s}\Phi(\v q,\v s)|\v q, \v s\rangle
\end{equation}
%with fixed total transverse  momentum $\v P$. 
We ignore the longitudinal momentum dependence for now 
(also the Lorentz and color indices).
In general we have
\begin{equation}\label{apr}
\langle \Phi|A^\dagger(\v k -\v p)A^\dagger(\v l+\v p)A(\v l)A(\v k)|\Phi\rangle =\Phi^*(\v k-\v p,\v l+\v p)\Phi(\v k,\v l)
\end{equation}
We now consider such states which in coordinate space have  size $R\sim1/\Lambda_{QCD}$, that is the coordinates of both particles are restricted to area $\v x^2,\,\v y^2<R^2\sim S_\perp$.
The wave function for such a state can be modeled in a factorized form  as
\begin{eqnarray}
\Phi(\v k, \v l)&=&\int _{\v x^2<R^2,\ \v y^2<R^2}e^{i\v k\cdot\v x+i\v l\cdot\v y}\chi_1(\v x+\v y)\chi_2(\v x-\v y)\\
&
\approx&\int_{|\v x+\v y|<R}e^{\frac{i}{2}(\v k+\v l)\cdot(\v x+\v y)}\chi_1(\v x+\v y)\int_{|\v x-\v y|<R}e^{\frac{i}{2}(\v k-\v l)\cdot(\v x-\v y)}\chi_2(\v x-\v y)\nonumber\\
&\approx&
\bar\chi_1(\v k+\v l) e^{i|\v k-\v l|R} \psi(\v k-\v l)\nonumber
\end{eqnarray}
where in the last approximate equality we have assumed that the integral over $\v x-\v y$ is dominated by the distances $\sim R$, so that $\psi(\v k -\v l)\approx const$ is a slowly varying function of momenta. This approximation in particular holds when the two gluons in their center of mass frame are localized within the distance of order $R\sim1/\Lambda_{QCD}$ of each other, while when the distance is smaller than $R$ the gluon positions are practically uncorrelated.
We then gat
\begin{equation}\label{wf}
\Phi^*(\v k-\v p,\v l+\v p)\Phi(\v k,\v l)\propto e^{iR\left(|\v k-\v l|-|\v k-\v l+2\v p|\right)}
\end{equation}
Since we are interested in values of momenta such that $|\v k|\sim|\v l|\sim |\v p|\gg \frac{1}{R}$, the phase in the above equation is oscillating fast and vanishes upon averaging over the small interval of momenta $\v p$ of order $1/R$, unless $ \v k=\v l+\v p, \ \ \v l=\v k-\v p$ with  accuracy $1/R$.

Thus the RHS of \eqref{apr} can be approximated by  \eqref{diluteaverage1} with the understanding that the $\delta$-function is regulated so that $\delta^2(0)\sim R^2=S_\perp$ with $R$ a phenomenological adjustable parameter with the physical meaning of the transverse size of the hadron.
Thus as far as the transverse momenta are concerned, \eqref{diluteaverage1} is qualitatively correct also for a state which contains superposition of momentum states, provided all momenta $\v l$, $\v k$ and $\v p$ are much larger than the nonperturbative scale associated with $1/R$. 

However the situation with the longitudinal momentum is somewhat different. The analog of the transverse size $R$ in the longitudinal direction is the longitudinal size of the "valence distribution" which is determined by the total longitudinal momentum of the hadron, $P^+$, i. e. $R_{longitudinal}\sim \frac{1}{P^+}$. On the other hand, the longitudinal momenta of gluons are always smaller than $P^+$. One is therefore never really in the situation similar to the transverse momenta case where the momenta measured by the observable are larger than the inverse size of the gluon distribution. Thus strictly speaking the delta function approximation is not suitable for the longitudinal momentum averaging in \eqref{diluteaverage1}. Nevertheless, if the gluon TMD is a slowly varying function of $k^+$, we can adjust this approximation to understand the basic physics of the nonlinear terms. We need to take into account the fact that if the momentum is small (smaller than the inverse size of the valence distribution) the actual size over which the gluons are distributed  in space is determined by their de Broiglie wave length. Therefore the role of $R$
in the analog of \eqref{apr} should be the inverse of the longitudinal momentum itself. Thus for our analysis of nonlinear terms we will use the following expression 
\begin{eqnarray}\label{diluteaverage}
&&\hspace{-2cm}\langle \Psi_P|: A^{\dagger b}_{l}(k^+-p^+,\v k-\v p)T^a_{bc}  A^c_{k}(k^+,\v k)
 A^{\dagger d}_{m}(l^++p^+,\v l+\v p)T^a_{de}  A^e_{n}(l^+,\v l):|\Psi_P\rangle_E\\
%&&=\frac{1}{4V(N_c^2-1)^2}\delta^3(k-l-p)\delta^{km}\delta^{ln}\delta^{cd}\delta^{be}T^a_{bc}T^a_{de}\langle\Psi_P|:\left[ A^{\dagger }(l^+,\v l) A(l^+,\v l)\right]\left[ A^\dagger(l^++p^+,\v l+\v p)
% A(l^++p^+,\v l+\v p) \right]:|\Psi_P\rangle\nonumber\\
&=&\frac{N_c}{4S_\perp(N_c^2-1)}\sqrt{l^+(l^++p^+)}\delta^3(k-l-p)\delta^{km}\delta^{ln}
\nonumber \\
&\times& \langle\Psi_P|:\left[ A^{\dagger }(l^+,\v l) A(l^+,\v l)\right]\left[ A^\dagger(l^++p^+,\v l+\v p)
 A(l^++p^+,\v l+\v p) \right]:|\Psi_P\rangle_E\nonumber
\end{eqnarray}
where in the prefactor we have taken a symmetric combination of $l^+$ and $k^+$ of correct dimensionality. 

We believe that the above expression is qualitatively correct for not very small values of $x$. For very small $x$ the analog of the phase factor in \eqref{wf} vanishes and the delta function approximation is not reasonable. In this limit it is better to keep the integral over the longitudinal momenta unrestricted. The price to pay then is that one would have to work with GPD's rather than TMD's, and we will not undertake this here. One does have to keep this caveat in mind, and be aware that our discussion of the nonlinear corrections below most likely is inadequate for very small values of $x$.

%In the longitudinal direction the size of the projectile state is given by the inverse of the total longitudinal momentum $P^+$, while the area in the transverse plane, $S$, should be though of as an effective parameter similar to the one introduced  in GLR equation, $V=S/P^+$. We will come back to this point later.
\subsubsection*{The stimulated emission}
In the approximation \eqref{diluteaverage}, equation \eqref{tmd2} becomes 
\begin{eqnarray}\label{oNL11e1}
&&\delta \mathcal{T}^{NL}(k)=\frac{g^2N_c}{(2\pi)^3}\frac{1}{4S_\perp(N_c^2-1)}\int\frac{d^3p}{(2\pi)^3}\frac{1}{2p^+}\frac{1}{2k^+}\\
&&\times\Bigg[
\frac{\sqrt{k^+(k^++p^+)}}{2(k+p)^+}F^i_{lk}(k+p,p)F^i_{lk}(k+p,p)\mathcal{T}_2(k,k+p)
-\frac{\sqrt{k^+(k^+-p^+)}}{2(k-p)^+}F^i_{lk}(k,p)F^i_{lk}(k,p)\mathcal{T}_2(k, k-p)\Bigg]\nonumber
\end{eqnarray}
where 
\begin{equation}
\mathcal{T}_2(k, q)\equiv \langle\Psi_P|:\left[a^{\dagger }(k^+,\v k)a(k^+,\v k)\right]\left[a^\dagger(q^+,\v q)
a(q^+,\v q) \right]:|\Psi_P\rangle_E
\end{equation}

The physical meaning of this is quite interesting. First, we note that $\mathcal{T}_2(k,q)$ is the number of {\it pairs} of gluons with momenta $k$ and $q$.
 Now suppose that the unevolved wave function contains a pair of gluons  with momenta $l$ and  $l+p$. In the step of the evolution the second gluon can emit a soft gluon with momentum $p$ so that now the wave function contains two gluons with momentum $l$. Counting these gluons (with the number operator) now is affected by the Bose statistics factor, since the two gluons are identical. This additional Bose contribution  is precisely the first, gain term in \eqref{oNL11e1}. 
This is nothing but  the stimulated emission correction which enhances the probability of transition $l+p\rightarrow l$ due to the presence of an additional gluon with momentum $l$ in the wave function.
The loss term  (the second term in \eqref{oNL11e1}) corresponds to the gluon $l$ emitting $p$ and therefore disappearing from the count. This contribution is the "stimulated emission" correction which enhances the transition rate  $l\rightarrow (p, l-p)$ gluon in the presence of an additional gluon with momentum $l-p$.

One naturally wonders whether the nonlinear correction \eqref{oNL11e1} is related to the nonlinear term in the famous  GLR  equation.
The answer is, it is not. The nonlinear term that appears in our calculation is due to the Bose statistics of  gluons and is of the same order in the coupling constant as the linear term. On the other hand the GLR, or Mueller-Qiu term is suppressed by an additional power of the coupling constant but enhanced by a factor $\ln x$. It arises due to the process where in a single step of the evolution two "valence" gluons from the wave function transition into one soft gluon and one "valence" gluon, or in the language of the BO evolution, two slow gluons merge into one slow gluon with the emission of one fast gluon. We do not consider such processes in the present work, as formally they are of the higher order in $\alpha_s$.

The stimulated emission term \eqref{oNL11e1} is proportional to the  density of pairs $\mathcal{T}_2$ which in principle is sensitive to correlations between gluons in the wave function. In the simplest case, when the gluons are uncorrelated, the density of pairs factorizes into product of single gluon densities. 
  One does expect this factorization to hold in the large $N_c$ limit, and in the rest of this paper for simplicity we will adhere to the factorization hypothesis
  \begin{equation}\label{fact}
  \mathcal{T}_2(k,q)\approx\mathcal{T}(k)\mathcal{T}(q)
  \end{equation}
  This simplifies things conceptually, since under this assumption the gluon TMD satisfies a closed albeit nonlinear equation\footnote{In general though one has to keep in mind that 
  if gluons are strongly correlated in the wave function the factorizability assumption breaks down and one has to treat this term more carefully.}. In the factorized approximation \eqref{fact}, the stimulated emission term is proportional to the gluon phase space density $\frac{1}{S}\mathcal{T}(\v k, x)$. This is a dimensionless factor which is small for a dilute system and plays the role of the suppression factor at large transverse momentum $\v k$.
  
  Combining the nonlinear contribution with the linear one we can write the evolution of the gluon TMD as
  \begin{eqnarray}\label{fulleqtmd}
\frac{\partial}{\partial \eta}\mathcal{T}(\v k, x)&=&-\frac{g^2N_c}{4\pi^2}\int_{0}^{1/2}d\zeta\left[\frac{\zeta}{1-\zeta}+\frac{1-\zeta}{\zeta}+\zeta(1-\zeta)\right] \nonumber \\
&\times&\Bigg[\mathcal{T}(\v k, x)\left\{1+\frac{x\sqrt{1-\zeta}}{16\pi^2S_\perp(N_c^2-1)}\int d\v n\ \mathcal{T}(\v k_{\zeta'},(1-\zeta)x)\right\} \nonumber \\
&-&
\frac{1}{1-\zeta}\int d\v n\ \mathcal{T}(\v k_\zeta,\frac{x}{1-\zeta})\left\{1+\frac{x}{16\pi^2S_\perp(N_c^2-1)\sqrt{1-\zeta}}\mathcal{T}(\v k,x)\right\}\Bigg]
 \end{eqnarray}
with 
\begin{eqnarray}&&\v k_{\zeta}=\v k+\v n\left[2k^+E\frac{\zeta}{1-\zeta}\right]^{1/2}=\v k+\v n\left[\v k^2\frac{E}{k^-}\frac{\zeta}{1-\zeta}\right]^{1/2}; \\
&&\v k_{\zeta'}=\v k+\v n\left[2k^+E\zeta\right]^{1/2}=\v k+\v n\left[\v k^2\frac{E}{k^-}\zeta\right]^{1/2}\nonumber\,.
\end{eqnarray}
As with the linear terms, the singularity at $\zeta=0$ cancels between the two induced emission contributions. 
%However for the induced emission terms the range of $\zeta$ where the cancellation occurs is different. Recall that for the linear terms, if $k^-\ll E$ the cancellation  takes place only in the range $\zeta\ll\frac{k^-}{E}$, whereas in the rest of the range of $\zeta$ one can neglect the gain term. For the induced emission terms on the other hand the cancellation happens for $\zeta\ll 1$ even if $k^-\ll E$.

For $\zeta\ll\frac{k^-}{E}$ we have $\v k_\zeta\approx\v k_{\zeta'}\approx\v k$ and thus there is a complete cancellation between the two nonlinear terms. To understand the behavior for larger $\zeta$ we will rely on the leading order perturbative behavior of the TMD, $\mathcal{T}(\v q,q^+)\propto \frac{1}{\v q^2}\frac{1}{q^+}$. With this form for $\zeta>\frac{k^-}{E}$ , we have (modulo perturbative corrections) 
\begin{eqnarray}\label{sub}
&&(1-\zeta)^{1/2}\mathcal{T}(\v k_{\zeta'},(1-\zeta)x)-\frac{1}{(1-\zeta)^{3/2}} \mathcal{T}(\v k_\zeta,\frac{x}{1-\zeta})
\\
&&\approx \frac{1}{(1-\zeta)^{1/2}}\mathcal{T}\left(\v n\left[\v k^2\frac{E}{k^-}\zeta\right]^{1/2},x\right)
-(1-\zeta)^{1/2}\mathcal{T}\left(\v n\left[\v k^2\frac{E}{k^-}\zeta\right]^{1/2},x\right)
\approx\zeta \mathcal{T}\left(\v n\left[\v k^2\frac{E}{k^-}\zeta\right]^{1/2},x\right)\nonumber
\end{eqnarray}
  In the last equality we have assumed $\zeta\ll 1$. This is justified since even with the explicit factor $\zeta$ in \eqref{sub}, the integral over $\zeta$ is dominated by small values of $\zeta$ due to the behavior of the TMD. With this assumption we have for the nonlinear term in \eqref{fulleqtmd}
  \begin{eqnarray}
&& \hspace{-2cm}-\frac{g^2N_c}{4\pi^2}\int_{0}^{1/2}d\zeta\left[\frac{\zeta}{1-\zeta}+\frac{1-\zeta}{\zeta}+\zeta(1-\zeta)\right]\mathcal{T}(\v k, x)\nonumber\\
&\times&\frac{x}{16\pi^2S_\perp(N_c^2-1)}\int d\v n\ \left\{(1-\zeta)^{1/2}\mathcal{T}(\v k_{\zeta'},(1-\zeta)x)
-\frac{1}{(1-\zeta)^{3/2}}\int \  \mathcal{T}(\v k_\zeta,\frac{x}{1-\zeta})
\right\}\nonumber\\
&=& -\frac{g^2N_c}{4\pi^2}\frac{x}{16\pi^2S_\perp(N_c^2-1)}\mathcal{T}(\v k, x)\int_{\v k^-/E}^{1/2}d\zeta\ \int d\v n \mathcal{T}\left(\v n\left[\v k^2\frac{E}{k^-}\zeta\right]^{1/2},x\right)\nonumber\\
&=& -\frac{g^2N_c}{4\pi^2}\frac{x}{4\pi S_\perp(N_c^2-1)}\frac{1}{\v k^2}\frac{k^-}{E}\mathcal{T}(\v k, x)\int_{\v p^2=\v k^2}^{\v p^2=k^+E} \frac{d^2\v p}{(2\pi)^2}\  \mathcal{T}\left(\v p,x\right)
 \end{eqnarray}

The TMD evolution equation including the nonlinear term therefore takes the form
\begin{eqnarray}
\frac{\partial}{\partial \eta}\mathcal{T}(\v k, x)&=&-\frac{g^2N_c}{4\pi^2}\Bigg[\int_{k^-/E}^{1/2}d\zeta\left[\frac{\zeta}{1-\zeta}+\frac{1-\zeta}{\zeta}+\zeta(1-\zeta)\right] \nonumber \\
&+&\frac{x}{4\pi S_\perp(N_c^2-1)}\frac{1}{\v k^2}\frac{k^-}{E}\int_{\v p^2=\v k^2}^{\v p^2=k^+E} \frac{d^2\v p}{(2\pi)^2}\  \mathcal{T}\left(\v p,x\right)\Bigg]
 \mathcal{T}(\v k, x)
\end{eqnarray}

Rewriting this in the more canonical form we obtain% In addition it is convenient to consider $x\mathcal{T}$ rather than $\mathcal {T}$ as a more fundamental  quantity. Written for this quantity the evolution equation becomes
  
\begin{eqnarray}\label{css+}
\frac{\partial}{\partial \eta}[x\mathcal{T}(\v k, x)]=-\frac{g^2N_c}{4\pi^2}\Bigg[&&\int_{\v k^2/Q^2}^{1/2}d\zeta \left[\frac{\zeta}{1-\zeta}+\frac{1-\zeta}{\zeta}+\zeta(1-\zeta)\right] \\
&+&\frac{1}{4\pi (N_c^2-1)}\frac{1}{Q^2S_\perp}\int_{\v p^2=\v k^2}^{\v p^2=Q^2} \frac{d^2\v p}{(2\pi)^2}\  [x\mathcal{T}\left(\v p,x\right)]\Bigg]
[x \mathcal{T}(\v k, x)]\nonumber
\end{eqnarray}
where we have introduced the transverse resolution scale $Q^2\equiv 2k^+E$ in accordance with our discussion in the previous subsection. We will also see below that this is the transverse resolution naturally arising in the context of the DGLAP equation for the gluon PDF.

We conclude with several comments about the nonlinear correction. First, as dictated by its origin in the Bose statistics of gluons, it is suppressed by the factor $1/(N_c^2-1)$. Second, at high $Q^2$ it is suppressed by the factor $1/S_\perp Q^2$, where $S_\perp$ is the effective gluonic transverse area of the projectile. This is the same suppression factor at high $Q^2$ that arises in the GLR equation \cite{GLR,MQ}, even though the stimulated emission term has a very different nature from the saturation correction of GLR. The sign of the nonlinear term in \eqref{css+} is the same as that of the linear one, and so it speeds up the decay of the TMD by enhancing the rate of Sudakov emissions.  Finally we note that the lower limit of integration $\v p^2=\v k^2$ in \eqref{css+} has a natural interpretation. The momentum $\v p$ is the momentum of the gluon which is a product of splitting of the gluon $\v k$. The condition $\v p^2>\v k^2$ simply ensures that this splittings is in the DGLAP kinematics. Thus both in the linear term and in the stimulated emission correction the contributions from non-DGLAP splittings cancel between the loss and the gain terms.

\subsection{The "real term": the gluon TMD at $k^-=E$.}
So far we have analyzed the evolution of the TMD when $k^-<E$. 
To complete the discussion we will now  consider the case when the frequency of the tagged gluon is equal to the frequency $E$. This situation is not described by the CSS equation, but rather provides the initial condition to the CSS or CSS-BO equation \eqref{linear} at the resolution scale which is equal to the momentum in the TMD. Additionally it is an important ingredient in the DGLAP equation, which we will recover later in the BO approach.

The gluon number operator $\hat T$ at $k^-=E$ depends only on gluonic degrees of freedom in the window of the phase space opened by the evolution when the evolution parameter $E$ reaches the frequency $k^-$.
The evolution of any such variable is given by the following general expression (assuming $\langle 0|\hat O|0\rangle_F=0$)
\beq\label{nonL}
\delta\langle \hat O\rangle=\int_{p,\bar p:\,p^-=\bar p^-=E} \langle\Psi_P| \,C^{\dagger a}_i(p)\, \langle 0|A^a_i(p)\hat OA^{\dagger b}_j(\bar p)|0\rangle_F\, C^b_j(\bar p)\,|\Psi_P\rangle_E
\eeq
where the inner average is taken over the vacuum of the fast modes in the window and the outer in the state of the slow modes evolved up to frequency $E$. 
For the TMD this gives
\beq
\delta \mathcal{T}(\v k,k^+; k^-=E)=\frac{1}{2k^+} \langle\Psi_P| C^{\dagger a}_i(k)C^a_i( k)|\Psi_P\rangle_E
\eeq
Writing this in the normal ordered form and using the dilute approximation for the nonlinear term we find
\begin{eqnarray}\label{realtmd}
\delta \mathcal{T}(\v k,k^+; k^-=E)&=&\frac{g^2N_c}{4\pi^3}\frac{1}{\v k^2}\int d^2\v q\int^{min\left(\frac{\v k^2}{\v q^2},\frac{\v k^2}{\v k^2+(\v q-\v k)^2}\right)}_{x} d\zeta \frac{1}{\zeta}\left[\frac{\zeta}{1-\zeta}+\frac{1-\zeta}{\zeta}+\zeta(1-\zeta)\right]\nonumber\\
&&\times\mathcal{T}(\v q, \frac{k^+}{\zeta})\Big[1+\frac{1}{16\pi^2S_\perp(N_c^2-1)}\frac{k^+(1-\zeta)^{1/2}}{\zeta}\mathcal{T}(\v q-\v k,k^+\frac{1-\zeta}{\zeta})\Big]
%&=&\frac{1}{2}\frac{g^2}{(2\pi)^3}\frac{1}{\v k^2}\frac{1}{2\pi}\frac{1}{4}\frac{1}{(2\pi)^3}32\int^{\v k^2=2k^+E}\frac{d^2q}{(2\pi)^2}\int^{1/2}_{x} d\zeta \frac{1}{\zeta}\left[\frac{\zeta}{1-\zeta}+\frac{1-\zeta}{\zeta}+\zeta(1-\zeta)\right]\nonumber\\
%&&\times\mathcal{T}(\v q, \frac{k^+}{\zeta})\Big[1+\frac{2}{4V(N_c^2-1)}\mathcal{T}((\v q-\v k)^2,k^+\frac{1-\zeta}{\zeta})\Big]+\delta^{BFKL}\mathcal{T}
\end{eqnarray}
%where $x\equiv \frac{k^+}{P^+}$. 
The integration limits on $\zeta$ follow from the conditions $k^->q^-$ and $k^->(k-q)^-$.
An equivalent form is
\begin{eqnarray}\label{realtmds}
\delta \left[x\mathcal{T}(\v k,x; k^-=E)\right]&=&\frac{g^2N_c}{4\pi^3}\frac{1}{\v k^2}\int d^2\v q\int^{min\left(\frac{\v k^2}{\v q^2},\frac{\v k^2}{\v k^2+(\v q-\v k)^2}\right)}_{x} d\zeta\left[\frac{\zeta}{1-\zeta}+\frac{1-\zeta}{\zeta}+\zeta(1-\zeta)\right]\\
&&\times\left[ \frac{x}{\zeta}\mathcal{T}(\v q, \frac{x}{\zeta})\right]\Bigg[1+\frac{1}{16\pi^2S_\perp(N_c^2-1)}\frac{1}{(1-\zeta)^{1/2}}\left[x\frac{1-\zeta}{\zeta}\mathcal{T}(\v q-\v k,x\frac{1-\zeta}{\zeta})\right]\Bigg]\nonumber
%&=&\frac{1}{2}\frac{g^2}{(2\pi)^3}\frac{1}{\v k^2}\frac{1}{2\pi}\frac{1}{4}\frac{1}{(2\pi)^3}32\int^{\v k^2=2k^+E}\frac{d^2q}{(2\pi)^2}\int^{1/2}_{x} d\zeta \frac{1}{\zeta}\left[\frac{\zeta}{1-\zeta}+\frac{1-\zeta}{\zeta}+\zeta(1-\zeta)\right]\nonumber\\
%&&\times\mathcal{T}(\v q, \frac{k^+}{\zeta})\Big[1+\frac{2}{4V(N_c^2-1)}\mathcal{T}((\v q-\v k)^2,k^+\frac{1-\zeta}{\zeta})\Big]+\delta^{BFKL}\mathcal{T}
\end{eqnarray}
The nonlinear term has the same meaning as in the previous subsection. It is  the induced emission correction to the splitting $(q\rightarrow k, q-k)$ due to the presence of additional gluons with momentum $q-k$ prior to the splitting.

Note that if we restrict ourselves to the DGLAP kinematics, i.e. $\v q^2\ll \v k^2$, the integration over the longitudinal momentum fraction $\zeta$ is limited by $\zeta<1/2$. This is the consequence of the fact that the gluon $k$ has the highest frequency, and therefore in DGLAP kinematics its longitudinal momentum is smaller than that of its partner, $k^+<(q-k)^+$.

In addition to the DGLAP splittings, \eqref{realtmd} contains contributions  from non-DGLAP splittings  with $\v q^2=(\v q-\v k)^2\gg \v k^2$. This region has a very limited longitudinal phase space  $\zeta<\v k^2/\v q^2$ and is unimportant for $x$ not too small. However for very small $x$ it can be significant.
It is instructive therefore to write \eqref{realtmds} as the contribution from two distinct kinematic regions
\begin{equation}
\delta \left[x\mathcal{T}(\v k,x; k^-=E)\right]=\delta \left[x\mathcal{T}(\v k,x; k^-=E)\right]^{DGLAP}+\delta \left[x\mathcal{T}(\v k,x; k^-=E)\right]^{BFKL}
\end{equation}
with
\begin{eqnarray}\label{realtmddglap}
&&\delta \left[x\mathcal{T}(\v k,x; k^-=E)\right]^{DGLAP}=\frac{g^2N_c}{4\pi^3}\frac{1}{\v k^2}\int^{1/2}_{x} d\zeta\left[\frac{\zeta}{1-\zeta}+\frac{1-\zeta}{\zeta}+\zeta(1-\zeta)\right]\\
&&\quad\times\int^{\v q^2<\v k^2} d^2\v q\left[ \frac{x}{\zeta}\mathcal{T}(\v q, \frac{x}{\zeta})\right]\Bigg[1+\frac{1}{16\pi^2S_\perp(N_c^2-1)}\frac{1}{(1-\zeta)^{1/2}}\left[x\frac{1-\zeta}{\zeta}\mathcal{T}(\v k,x\frac{1-\zeta}{\zeta})\right]\Bigg]\nonumber
\end{eqnarray}
and
\begin{eqnarray}\label{realtmdbfkl}
\delta \left[x\mathcal{T}(\v k,x; k^-=E)\right]^{BFKL}&=&\frac{g^2N_c}{4\pi^3}\frac{1}{\v k^2}\int^{1}_{x} d\zeta\frac{1}{\zeta}\int_{\v q^2=\v k^2}^{\v q^2=\frac{\v k^2}{\zeta}} d^2\v q\left[ \frac{x}{\zeta}\mathcal{T}(\v q, \frac{x}{\zeta})\right]
%\Bigg[1+\frac{1}{16\pi^2S(N_c^2-1)}\frac{\zeta}{x}\left[\frac{x}{\zeta}\mathcal{T}(\v q,\frac{x}{\zeta})\right]\Bigg]\nonumber
\end{eqnarray}
where we have assumed that $\zeta\ll 1$ in \eqref{realtmdbfkl}. 
The BFKL contribution \eqref{realtmdbfkl} is only sizable for $x\ll 1$, since only in this case one has a large enough phase space for the $\v q$ integration.

 Note that we did not include the nonlinear term in the BFKL contribution. It is easy to see that at $\v q^2>\v k^2$ the high $\v q$ part of the integrand in the nonlinear term  is proportional to $\mathcal{T}^2(\v q)$, and the integral over $\v q$ is therefore dominated by the lower limit, $\v q^2\sim \v k^2$. This range is already taken into account in \eqref{realtmddglap} and so there is no need to include it in \eqref{realtmdbfkl}.
 
 As noted above, at very small $x$ the contribution of \eqref{realtmdbfkl} is large and in principle comparable with that of \eqref{realtmddglap}. In particular for a simple perturbative form $\mathcal{T}(\v q)\sim 1/\v q^2$, the transverse integral in \eqref{realtmdbfkl} yields $\ln 1/\zeta$, and the subsequent integral over $\zeta$ gives $\ln^2 1/x$.
On the other hand \eqref{realtmddglap} parametrically yields $\ln 1/x\ln\frac{\v k^2}{\Lambda^2_{QCD}}$. Thus both contributions are doubly logarithmic, and which one is dominant depends on the values of $x$ and $\v k^2$. We conclude that for small $x$ the contribution from the BFKL kinematics, which does not originate from  the doubly logarithmic region is large and needs to be retained.

%\begin{eqnarray}\label{tmdbf}
%\delta^{BFKL}\mathcal{T}&=&\frac{1}{2}\frac{g^2}{(2\pi)^3}\frac{1}{\v k^2}\frac{1}{2\pi}\frac{1}{4}\frac{1}{(2\pi)^3}32\int^{1/2}_{x} d\zeta \frac{1}{\zeta}\left[\frac{\zeta}{1-\zeta}+\frac{1-\zeta}{\zeta}+\zeta(1-\zeta)\right]\nonumber\\
%&&\times\int_{\v k^2}^{\v k^2/\zeta}\frac{d^2q}{(2\pi)^2}\mathcal{T}(\v q, \frac{k^+}{\zeta})\Big[1+\frac{2}{4V(N_c^2-1)}\mathcal{T}((\v q-\v k)^2,k^+\frac{1-\zeta}{\zeta})\Big]
%\end{eqnarray}

%If $x$ is not too small the phase space for transverse integration in $\delta^{BFKL}\mathcal{T}$ vanishes and this term is negligible. For $x\ll 1$ the term $\delta^{BFKL}\mathcal{T}$ is not negligible. This additional term originates from splittings which are outside of the standard DGLAP kinematics: $q\rightarrow (k, q-k)$ with $q^+\gg k^+$, $\v q^2\gg \v k^2$ ( for very small $x$).  These are close to eikonal or eikonal splittings which reflect BFKL physics, hence our notation in \eqref{realtmd},\eqref{tmdbf}.

One notices that there is no explicit factor of $\Delta$ on the RHS of this equation. The reason is clear - this contribution exists only when the evolution parameter $\eta$ passes through the value $\ln k^-/E_0$. This is therefore a delta function contribution to the evolution equation and the result \eqref{realtmd} should be understood as the integrated change in the value of TMD when $\eta$ changes from $\ln k^-/E_0-\Delta/2$ to $\ln k^-/E_0+\Delta/2$. Thus as a contribution to the evolution equation this can be written as
\begin{eqnarray}\label{realtmd1}
&&\frac{\partial}{\partial\eta}\left[x \mathcal{T}_{real}^{DGLAP}(\v k,x)\right]=
\frac{g^2N_c}{4\pi^3}\frac{1}{\v k^2}\int^{1/2}_{x} d\zeta\left[\frac{\zeta}{1-\zeta}+\frac{1-\zeta}{\zeta}+\zeta(1-\zeta)\right]\\
&&\quad \times\int^{\v q^2<\v k^2} d^2\v q\left[ \frac{x}{\zeta}\mathcal{T}(\v q, \frac{x}{\zeta})\right]\Bigg[1+\frac{1}{16\pi^2S_\perp(N_c^2-1)}\frac{1}{(1-\zeta)^{1/2}}\left[x\frac{1-\zeta}{\zeta}\mathcal{T}(\v k,x\frac{1-\zeta}{\zeta})\right]\Bigg]\nonumber
%\frac{g^2N_c}{4\pi^3}\frac{1}{\v k^2}\int d^2\v q\int^{min\left(\frac{\v k^2}{\v q^2},\frac{\v k^2}{\v k^2+(\v q-\v k)^2}\right)}_{x} d\zeta \frac{1}{\zeta}\left[\frac{\zeta}{1-\zeta}+\frac{1-\zeta}{\zeta}+\zeta(1-\zeta)\right]\nonumber\\
%&&\times\mathcal{T}(\v q, \frac{k^+}{\zeta})\Big[1+\frac{1}{16\pi^2V(N_c^2-1)}\mathcal{T}((\v q-\v k)^2,k^+\frac{1-\zeta}{\zeta})\Big]
\delta(\eta-\ln \frac{k^-}{E_0})
\end{eqnarray}
\begin{eqnarray}
\frac{\partial}{\partial\eta} \left[x\mathcal{T}_{real}^{BFKL}(\v k,x)\right]&=&\frac{g^2N_c}{4\pi^3}\frac{1}{\v k^2}\int^{1}_{x} d\zeta\frac{1}{\zeta}\int_{\v q^2=\v k^2}^{\v q^2=\frac{\v k^2}{\zeta}} d^2\v q\left[ \frac{x}{\zeta}\mathcal{T}(\v q, \frac{x}{\zeta})\right]\delta(\eta-\ln \frac{k^-}{E_0})
\end{eqnarray}

At not too small values of  $x$ the linear term in \eqref{realtmddglap} is strongly reminiscent of the usual perturbative relation between PDF and the TMD at highest available transverse momenta. However there is a peculiarity here. Since at $\v k^2=2k^+E$ the  particles in the TMD have the highest frequency present in the wave function, at this value of $\v k$ the BO cascade does not contain gluons with longitudinal momentum lower than $k^+$. For this reason the integral over $\zeta$ in \eqref{realtmd} is cutoff  from above by $1/2$ and not by $1$ as in the usual DGLAP/CSS cascade. As we will see in the next subsection, when translated into evolution of the PDF, the  contribution from the "missing" range of the integration over $\zeta$ is actually contained in the gain term contributing to the second term in \eqref{realtmd1}.

This concludes our discussion of the BO evolution of the TMD at fixed transverse and longitudinal momentum. 
We next turn to the evolution of the gluon PDF.

\section{The PDF and the DGLAP equation}
The natural way to define the  gluon PDF in the context of the BO evolution is
\beq\label{pdfm}
G( x,E)\equiv\int d^2\v k\mathcal{T}(\v k,k^+;E)=\int_0^{\v 2k^+E}d^2\v k\mathcal{T}(\v k,k^+;E)\ ,
\eeq
as this quantity is proportional to the scattering cross section off a target with high transverse momentum transfer 
\eqref{dis}.
%\eqref{scat}, \eqref{hscat}.

$G(x,E)$ in \eqref{pdfm}  counts the total number of gluons with longitudinal momentum $k^+$ in the wave function evolved to frequency $E$. If compared to the standard definition of the PDF expressed in terms of the integral of the TMD, \eqref{pdfm} %suggests that we should view the transverse resolution scale in this definition as 
implies the transverse resolution scale is
\beq Q^2=2k^+E=2xs\,,\qquad\qquad x=k^+/P^+.
\eeq 
%where $s$ is the square of the total energy of the process (in the center of mass frame). In the last equality we have used the fact that the minus component of the target momentum in our calculation $q^-\approx E$, and $k^+=xP^+$. 
At small $x$ this is the standard relation between the Bjorken $x$, momentum transfer and the energy in DIS.

This relation is consistent with our discussion of the CSS equation for the choice of the parameter $a=\sqrt{2}$ in \eqref{resol}, which as we have explained above is also the preferred value of $a$ from the point of view of the applications of the CSS equation to physical processes.

The evolution of the PDF with frequency is of course not independent from the evolution of the TMD, but is given by
\begin{eqnarray}\label{evpdf}
\frac{\partial}{\partial \ln E}G(x,E)&=&(2k^+E)\pi\mathcal{T}(\v k^2=2k^+E,k^+;E)+\int_0^{2k^+E}d^2\v k \frac{\partial}{\partial \ln E}\mathcal{T}(\v k,k^+;E)
\end{eqnarray}
The first term here is given by the real term in the evolution of  the TMD, while the second term is given by the Lindblad term, both discussed in the previous section.

In this section we demonstrate that in the leading logarithmic approximation, the linear terms in the evolution of the gluon PDF in \eqref{evpdf} combine into the DGLAP equation as long as $x$ is not too small. For very small $x$, Eq.\eqref{evpdf} differs from the DGLAP equation by a term which arises from the BFKL kinematics of the splittings.

\subsection{Linear evolution at moderate $x$}

We start by considering  the second term in \eqref{evpdf} (in this section we only consider the linear terms in the evolution). To calculate it we could directly use \eqref{linear}. However it is more convenient to represent this contribution a little differently.

Consider \eqref{tmd1e}, but now instead of integrating explicitly over $\v p^2$, let us integrate over $p^+$ first.
We will analyze the gain and the loss terms separately. 

\subsubsection{The gain term}

Let us deal with  the gain term in \eqref{tmd1e} first. Integrating over $p^+$ utilizing the constraint $p^-=E$, as in \eqref{intwin}, it can be written as
\begin{eqnarray}\label{gain2}
&&\frac{g^2N_c}{2}\int \frac{d^3p}{(2\pi)^3}\frac{1}{2p^+}\frac{1}{4k^+(k^++p^+)}F^l_{st}(k+p,p)F^l_{st}(k+p,p)\mathcal{T}(\v k+\v p,k^++p^+)
\\
&&=\frac{g^2N_c}{2}\frac{\Delta}{\pi}\frac{1}{Ek^+}\int \frac{d^2\v p}{(2\pi)^2}\bar\zeta\left[\frac{\bar\zeta}{1-\bar\zeta}+\frac{1-\bar\zeta}{\bar\zeta}+\bar\zeta(1-\bar\zeta)\right]\mathcal{T}(\v k+\v p,\frac{x}{\bar\zeta})\nonumber
\end{eqnarray}
where \beq
\bar\zeta\equiv\frac{k^+}{k^++p^+}=\frac{1}{1+\frac{\v p^2}{2Ek^+}}
\eeq
To calculate the change in the PDF we have to integrate this over $\v k$. This is conveniently done by changing variables from $\v k$ to $\v q=\v k+\v p$, and from $\v p^2$ to $\bar \zeta$ (we will drop the bar from $\zeta$ for simplicity of notation).
\begin{eqnarray}\label{gainl}
&&\frac{g^2N_c}{2}\int d^2\v k\frac{\Delta}{\pi}\frac{1}{Ek^+}\int \frac{d^2\v p}{(2\pi)^2}\bar\zeta\left[\frac{\bar\zeta}{1-\bar\zeta}+\frac{1-\bar\zeta}{\bar\zeta}+\bar\zeta(1-\bar\zeta)\right]\mathcal{T}(\v k+\v p,\frac{x}{\bar\zeta})\\
&&=\frac{g^2N_c}{2}\frac{\Delta}{2\pi^2}\int d\zeta\left[\frac{\zeta}{1-\zeta}+\frac{1-\zeta}{\zeta}+\zeta(1-\zeta)\right]\int_{\Omega} d^2\v q\frac{1}{\zeta}\mathcal{T}(\v q,\frac{x}{\zeta})\nonumber
\end{eqnarray}
The integration region $\Omega$ in \eqref{gainl} is determined by the two basic constraints:
\beq\label{constra}
q^-<E; \ \ \ \ \ \ \ (q-p)^-<E.
\eeq
The integrals in \eqref{gainl} are over the two dimensional vector $\v q$ and the momentum fraction $\zeta$. The integral over the angle of $\v q$ in principle is not trivial, even though the TMD does not depend on this angle, since the value of the angle affects the limits of the integration over $\v q^2$ and $\zeta$. The first constraint in \eqref{constra} does not need to be imposed explicitly since it is automatically enforced by the presence of the TMD at momentum $\v q$ which vanishes outside the pertinent limits. Still, it is useful to write it out
\beq\label{c11}
\v q^2<\frac{2Ek^+}{\zeta}; \ \ \ \ \ \ \rm{or} \ \ \ \ \ \zeta<\frac{2Ek^+}{\v q^2}
\eeq
The second constraint can be written as
\beq \label{constrai}
\v q^2+\v p^2-2|\v p||\v q|\cos\phi<2Ek^+; \ \ \ \ \ \ \ \frac {\v q^2}{2Ek^+}-2\sqrt{\frac{\v q^2}{2Ek^+}}\sqrt{\frac{1-\zeta}{\zeta}}\cos\phi<\frac{2\zeta-1}{\zeta}
\eeq
where $\phi$ is the angle between $\v q$ and $\v p$.
To understand the effect of this constraint, first assume that the splittings are in the strict DGLAP kinematics, i.e. $q^2\ll 2Ek^+\frac{1-\zeta}{\zeta}$. Dropping the terms involving the small ratio we find that the constraint does not limit the integration range over the angle $\phi$, while for $\zeta$ it gives
\beq\label{121}
\frac{1}{2}<\zeta<1
\eeq
This is exactly the range of $\zeta$ integration which is missing in \eqref{realtmd} in order to complete the real contribution to the RHS in \eqref{evpdf} to the full range that contributes to the standard DGLAP equation.

Following the usual logic invoked in the derivation of the DGLAP evolution we can then argue that expansion of \eqref{constrai} in powers of $\v q^2/2Ek^+$ leads in \eqref{gainl} to terms which are suppressed at least by the factor $\log Q^2/\Lambda^2_{QCD}$, or equivalently $\alpha_s(Q^2)$.
Imagine solving the constraint \eqref{constrai} for $\zeta$ and expanding the solution in powers of $\v q^2/2Ek^+$:
\beq
\frac{1}{2}+\beta_-(\cos\phi, \frac{\v q^2}{2Ek^+})<\zeta<1+\beta_+(\cos\phi, \frac{\v q^2}{2Ek^+})
\eeq
where $\beta_-$ and $\beta_+$ have power expansion. Integration over $\zeta$ and $\phi$ in \eqref{gainl} then produces corrections in powers of $\v q^2/2Ek^+$ calculated with the limits \eqref{121}. Subsequently one should integrate over $\v q$. In perturbation theory the TMD behaves as $T(\v q^2)\sim 1/\v q^2$ modulo logarithmic corrections. Thus the integral of the TMD yields a factor of $\ln (2Ek^+/\Lambda_{QCD}^2)$, while for the TMD multiplied by any positive power of $\v q^2$ the integral is not logarithmic and yields a constant. Thus to leading logarithmic accuracy one can disregard the functions $\beta_-$ and $\beta_+$, and simply integrate in the limits \eqref{121}.

If we consider the angular integration in \eqref{gainl} first, we see that the integration region over the angle covers the whole $2\pi$ range as long as
\beq\label{c1}
\frac{\v q^2}{2Ek^+}<\frac{1}{\zeta}-2\sqrt{\frac{1-\zeta}{\zeta}}
\eeq
On the other hand the angular integration region shrinks to zero when
\beq\label{c2}
\frac{\v q^2}{2Ek^+}>\frac{1}{\zeta}+2\sqrt{\frac{1-\zeta}{\zeta}}
\eeq
For $\zeta\ne 1$ the condition \eqref{c1} is more restrictive than \eqref{c11}. However as long as $\zeta>1/2$ the difference between \eqref{c1} and \eqref{c11} is not significant, and within the leading logarithmic approximation the two are equivalent. Moreover, \eqref{c2} shows that whenever \eqref{c11} is satisfied, the angular integration region is finite.  
Therefore the value of the integral in \eqref{gainl} can be obtained by taking $-\pi<\phi<\pi$ and setting the integration limit for the  $\v q^2$ integral at some point $Q^2$ between the values \eqref{c1} and \eqref{c11}. The precise value of $Q^2$ can in principle be determined, but it only affects corrections suppressed by $\ln \frac{Q^2}{\Lambda_{QCD}^2}$ as per the argument above.

One could worry about the close vicinity of the point $\zeta=1/2$. At $\zeta=1/2$ the constraint on the angle is nontrivial
\beq
\cos\phi>\frac{1}{2}\sqrt{\frac{\v q^2}{2Ek^+}}
\eeq
For small $\v q^2$ this restricts $\phi$ to the range $-\frac{\pi}{2}<\phi<\frac{\pi}{2}$, while for the largest possible value of $\v q$, i.e. $\v q^2=4Ek^+$ to $-\frac{\pi}{4}<\phi<\frac{\pi}{4}$, and thus the angular range is smaller than $2\pi$ for any allowed value of $\v q^2$. Nevertheless, since the splitting function is finite at $\zeta=1/2$,  the point $\zeta=1/2$ has zero measure in the $\zeta$ integral, and the previous argument is unaffected.

Thus with the logarithmic accuracy we can disregard the variation of the angular integration region with $\v q^2$ and $\zeta$, and take $-\pi<\phi<\pi$ for all values of $\v q^2$ and $\zeta$  given by \eqref{c1}. We conclude that  in the leading logarithmic approximation  the gain term is given by
\begin{eqnarray}\label{gainl1}
&&\frac{g^2N_c}{2}\int d^2\v k\frac{\Delta}{\pi}\frac{1}{Ek^+}\int \frac{d^2\v p}{(2\pi)^2}\bar\zeta^2\frac{1}{\bar\zeta}\left[\frac{\bar\zeta}{1-\bar\zeta}+\frac{1-\bar\zeta}{\bar\zeta}+\bar\zeta(1-\bar\zeta)\right]\mathcal{T}(\v k+\v p,\frac{x}{\bar\zeta})\nonumber\\
&&\approx\frac{g^2N_c}{2}\frac{\Delta}{2\pi^2}\int_{1/2}^{1} d\zeta\frac{1}{\zeta}\left[\frac{\zeta}{1-\zeta}+\frac{1-\zeta}{\zeta}+\zeta(1-\zeta)\right]\int_0^{2Ek^+/\zeta} d^2\v q\mathcal{T}(\v q,\frac{x}{\zeta})
\end{eqnarray}
At fixed $k^+$ and $Q^2=2k^+E$ we can therefore write the gain contribution to the evolution of the PDF as
\beq 
\frac{\partial}{\partial \ln Q^2}\left[xG(x,E)\right]^{gain}=\frac{g^2N_c}{4\pi^2}\int_{1/2}^{1} d\zeta\left[\frac{\zeta}{1-\zeta}+\frac{1-\zeta}{\zeta}+\zeta(1-\zeta)\right]\int_0^{Q^2/\zeta} d^2\v q\left[\frac{x}{\zeta}\mathcal{T}(\v q,\frac{x}{\zeta})\right]
\eeq

\subsubsection{The loss (virtual) term }
The loss term in \eqref{tmd1e} does not call for any special treatment. The only manipulation we do to get it into the standard form is to change the integration variable $\zeta\rightarrow 1-\zeta$ and take half the sum of the two expressions. The result is 
\begin{eqnarray}
&&-\frac{g^2N_c}{2}\int \frac{d^3p}{(2\pi)^3}\frac{1}{8p^+k^+(k^+-p^+)}F^l_{st}(k,p)F^l_{st}(k,p)\mathcal{T}(k)\nonumber \\
&&\qquad=-\Delta\frac{g^2N_c}{8\pi^2}\int_0^{1}d\zeta\left[\frac{\zeta}{1-\zeta}+\frac{1-\zeta}{\zeta}+\zeta(1-\zeta)\right]\mathcal{T}(\v k,x)
\end{eqnarray}
Thus the contribution of this term to the evolution of PDF \eqref{evpdf} is 
\beq
\frac{\partial}{\partial\ln Q^2}\left[xG(x,E)\right]^{loss}=-\frac{g^2N_c}{8\pi^2}\int_0^{1}d\zeta\left[\frac{\zeta}{1-\zeta}+\frac{1-\zeta}{\zeta}+\zeta(1-\zeta)\right]\int_0^{Q^2}d^2\v k\left[x\mathcal{T}(\v k,x)\right]
\eeq

\subsubsection{The "real" term}
The first term in \eqref{evpdf} is given directly by \eqref{realtmd}.  %We  now concentrate on the linear term in \eqref{realtmd}. 
The integration limits in \eqref{realtmd} are shaped by two constraints
\beq 
q^-<E; \ \ \ \ (q-k)^-<E
\eeq
Given that $\v k^2=2k^+E$ the two constraints become
\beq\label{max}
\v q^2<2E k^+/\zeta
\eeq
and
\beq
\frac{\v q^2}{2Ek^+}-2\sqrt{\frac{\v q^2}{2Ek^+}}\cos\phi<\frac{1-2\zeta}{\zeta}
\eeq
where $\phi$ is the angle between $\v q$ and $\v k$.
In the DGLAP kinematics, where $\v q^2\ll 2Ek^+$ the second constraint becomes
\beq\label{min}
\zeta<\frac{1}{2}
\eeq
The angular integration region covers the full $2\pi$ for 
\beq \frac{\v q^2}{2Ek^+}<\frac{1}{\zeta}-2\sqrt{\frac{1-\zeta}{\zeta}}
\eeq
while it shrinks to zero for
\beq \frac{\v q^2}{2Ek^+}>\frac{1}{\zeta}+2\sqrt{\frac{1-\zeta}{\zeta}}
\eeq
The same argument as following \eqref{constrai} establishes that we can simplify the integration limits to \eqref{max} and \eqref{min}:
\begin{equation}
Q^2\pi\mathcal{T}(\v k^2=Q^2,k^+,E)=\frac{g^2N_c}{4\pi^2}\int^{1/2 }_{x} d\zeta \left[\frac{\zeta}{1-\zeta}+\frac{1-\zeta}{\zeta}+\zeta(1-\zeta)\right]\int^{Q^2/\zeta}\frac{d^2\v q}{(2\pi)^2}\frac{1}{\zeta}\mathcal{T}(\v q, \frac{k^+}{\zeta})
\eeq
with $Q^2=2k^+E$.
Note that with this upper limit on the integration over $\v q$, this expression contains the linear contributions from both $\delta \mathcal{T}^{DGLAP}$ and $\delta\mathcal{T}^{BFKL}$ of \eqref{realtmddglap} and \eqref{realtmdbfkl}, and is therefore valid also for small values of $x$.

\subsubsection{The DGLAP equation}
We can now combine the above expressions for the individual contributions. We define the glue to glue splitting function in the standard way
\beq
P_{gg}(\zeta)=\left[\frac{\zeta}{1-\zeta}\right]_++\frac{1-\zeta}{\zeta}+\zeta(1-\zeta)
\eeq
with the usual $+$ prescription for the $\zeta=1$ pole. 
%We also define the resolution scale for the PDF as $Q^2\equiv 2Ek^+$, which is a natural definition in view of our discussion in the previous section. 
We can now write the linear contribution to \eqref{evpdf} in the leading logarithmic approximation as
\beq\label{dglap+}
\frac{\partial}{\partial \ln Q^2}\left[x G(x,Q^2)\right]=\frac{\alpha_s}{2\pi}\int_x^1d\zeta P_{gg}(\zeta)\left[\frac{x}{\zeta}G\left(\frac{x}{\zeta},\frac{Q^2}{\zeta}\right)\right]
\eeq
with $Q^2\equiv 2Ek^+$.
As we have discussed above, in the leading logarithmic approximation we can drop the factor $1/\zeta$ in the transverse resolution scale on the RHS as long as $\zeta$ is not very small, which is always true for moderate values of $x$.  In this approximation \eqref{dglap+} becomes the standard DGLAP equation. At small $x$ however there is a contribution from splittings in the BFKL kinematics, and their effect is to change $Q^2$ into $Q^2/\zeta$ on the right hand side of the DGLAP equation.

%\begin{figure}[t]         
%\centering                              
   %                               \includegraphics[width=6cm]{DGLAPvsBO PDF.pdf}             
%\caption{The difference between the DGLAP and BO cascades. In (A) both pairs of particles (denoted by crosses and triangles) are created for the first time by splittings in the last step of the evolution. Both splittings  contribute to the first term in \eqref{dg}. In (B) only the particles residing in the diagonal strip are created for the first time in the last step, while the particles in the other two points are emitted in addition to those already present in the wave function. The pair denoted by crosses contributes to the first term in \eqref{evpdf}, while the pair denoted by triangles contributes to the second term in \eqref{evpdf}.  }
%\label{f3}
%\end{figure}

\begin{figure}[t] % Use figure environment for captions
\centering % Center the plot
\begin{tikzpicture}
\begin{axis}[
        axis lines=left,             % Keep axes at the left and bottom
        axis line style={-latex},
	 xmin=0,   xmax=4,
	 ymin=0,   ymax=4,
         %extra x ticks={-1,1},
	 %extra y ticks={-2,2},
	 xtick=\empty,
	 ytick=\empty,
	 xlabel={$\ln k^2$}, % Optionally keep labels without numbers
         ylabel={$\ln k^+$},
          x label style={at={(axis description cs:1,0)}}, 
          y label style={rotate=270, at={(axis description cs:0,1)}},
	%extra tick style={grid=major}, 
	clip=false,
	]    
         \addplot [only marks,mark=*] coordinates { (1,2) };
	% add dashed line
	\addplot[dash pattern=on 5pt off 3pt] coordinates {(0,2) (4,2)};
	 \node at (axis description cs:-0.15,0.5) [anchor=west] {$x_{Bj}$};
	 \addplot[domain=0:4] ({2},{x}); % A vertical line at x=2
	 \addplot[domain=0:4] ({2.8},{x});
	 \node at (axis description cs:0.5,-0.15) [anchor=south] {$\ln Q^2$};
	 \node at (axis description cs:0.7,-0.15) [anchor=south] {$\qquad\Delta+ \ln Q^2 $};
	  \addplot[only marks, mark=x, mark size=4pt, mark options={thick, black}] coordinates {(2.1, 2) (2.1, 1)};
	  \addplot[only marks, mark=triangle*, mark size=4pt, mark options={thick, black}] coordinates {(2.5, 2) (2.5,3)};
	\end{axis}
\end{tikzpicture}
\caption{DGLAP-cascade, where both pairs of particles (denoted by crosses and triangles) are created for the first time by splittings in the last step of the evolution. Both splittings  contribute to the first term in \eqref{dg}.}
\label{Fig:DGLAP_cascade} % Optional: label for referencing the figure
\end{figure}
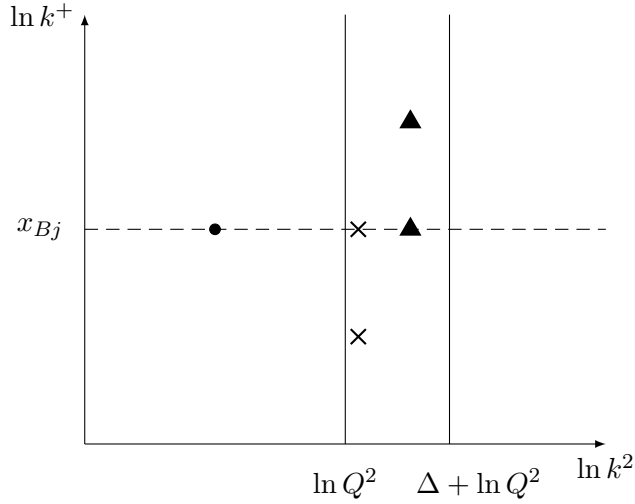

\begin{figure}[t] % Use figure environment for captions
\centering % Center the plot
\begin{tikzpicture}
\begin{axis}[
        axis lines=left,             % Keep axes at the left and bottom
        axis line style={-latex},
	 xmin=0,   xmax=6,
	 ymin=0,   ymax=6,
         %extra x ticks={-1,1},
	 %extra y ticks={-2,2},
	 xtick=\empty,
	 ytick=\empty,
	 xlabel={$\ln k^2$}, % Optionally keep labels without numbers
         ylabel={$\ln k^+$},
          x label style={at={(axis description cs:1,0)}}, 
          y label style={rotate=270, at={(axis description cs:0,1)}},
	%extra tick style={grid=major}, 
	clip=false,
	]    
         \addplot [only marks,mark=*] coordinates { (0.5,1.5) };
	% add dashed line
	\addplot[dash pattern=on 5pt off 3pt] coordinates {(0,1.5) (6,1.5)};
	 \node at (axis description cs:-0.15,0.25) [anchor=west] {$x_{Bj}$};
	 %\addplot[domain=0:4] ({2},{x}); % A vertical line at x=2
	 %\addplot[domain=0:4] ({2.8},{x});
	 %\node at (axis description cs:0.5,-0.15) [anchor=south] {$\ln Q^2$};
	% \node at (axis description cs:0.7,-0.15) [anchor=south] {$\Delta+ \ln Q^2 $};
	  % Draw a straight line from (0, 3) to (3, 0)
         \addplot[domain=0:3, samples=2] {-x + 3}; % Line with a slope of 1 and intercept 3
          \addplot[domain=0:4.5, samples=2] {-x + 4.5}; 
	  \addplot[only marks, mark=x, mark size=4pt, mark options={thick, black}] coordinates {(2, 1.5) (2,0.75)};
	  \addplot[only marks, mark=triangle*, mark size=4pt, mark options={thick, black}] coordinates {(1, 1.5) (1, 2.5)};
	   \node at (axis description cs: 0.5,-0.15) [anchor=south] {$\ln E$};
	 \node at (axis description cs: 0.75,-0.15) [anchor=south] {$\Delta+ \ln E $};
	\end{axis}
\end{tikzpicture}
\caption{BO-cascade, only the particles residing in the diagonal strip are created for the first time in the last step, while the particles in the other two points are emitted in addition to those already present in the wave function. The pair denoted by crosses contributes to the first term in \eqref{evpdf}, while the pair denoted by triangles contributes to the second term in \eqref{evpdf}.  }
\label{Fig:BO_cascade} % Optional: label for referencing the figure
\end{figure}
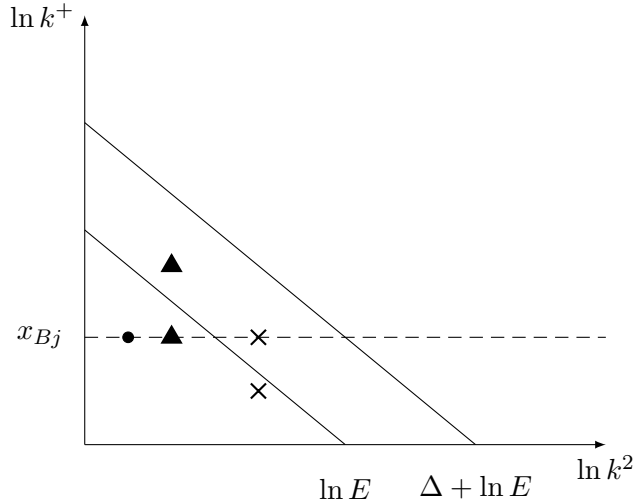

It is interesting that although the BO approach yields the standard DGLAP evolution at moderate $x$, the contributions to this equation are assembled somewhat differently compared to the usual DGLAP/CSS cascade.
Consider the relation between the PDF and the CSS TMD, the analog of \eqref{evpdf} in the DGLAP/CSS cascade
\beq\label{dg}
\frac{\partial}{\partial\ln Q^2}G(x,Q^2)=Q^2\pi{T}(\v k^2=Q^2,x; Q^2,\xi)+\int_k\frac{\partial}{\partial Q^2}{T}(\v k^2, x; Q^2,\xi)
\eeq
This equation is very similar to \eqref{evpdf}. However there are some differences between the two, since the standard DGLAP/CSS cascade and the BO cascade are not identical as discussed in the previous section.  

In the DGLAP/CSS  cascade in one step of the evolution one allows emission of higher $\v k$ gluons irrespective of their longitudinal momentum fraction. 
As a result, in the standard DGLAP/CSS cascade the complete real (gain) term in the DGLAP equation originates from the first term in \eqref{dg}: the TMD at the highest transverse momentum ($\v k^2=Q^2$) at any value of $x$ is generated by the last splitting in the cascade whereby the gluon with longitudinal momentum $k^+$ splits into two gluons with $\zeta k^+$ and $(1-\zeta)k^+$ with large transverse momenta and  any value of $x<\zeta<1$ (Fig. \ref{Fig:DGLAP_cascade}). Thus for a fixed value of $x$ the TMD is generated by splitting of gluons with $x<\zeta<1/2$ as well as $1/2<\zeta<1$. 

On the other hand the BO cascade allows only emission of particles with higher frequency.
  Thus in the last step of the BO evolution a new gluon is  only created by splittings with $\zeta<1/2$, as otherwise its  partner gluon would  be the one with highest frequency, Fig. \ref{Fig:BO_cascade}.  Therefore only part of the real contribution to the DGLAP equation originates from the first term in \eqref{evpdf}. The remaining contribution comes from the second term in \eqref{evpdf}, which counts the increase in the number of gluons that are already present in the wave function.

%  The difference between the origin of the real contributions in DGLAP and BO evolutions is demonstrated in Fig. \ref{Fig:BO_cascade}. 
The difference between the origin of the real contributions in DGLAP/CSS and BO cascades notwithstanding, the net result is that for not very small values of $x$ the two cascades in the leading logarithmic approximation lead to the very same evolution equation - the standard DGLAP.

However for very small $x$ such that $\ln x=O(1/\alpha_s)$ one cannot neglect the factor $1/\zeta$ in the transverse resolution scale on the RHS in \eqref{dglap+}, since the integral over small values of $\zeta$ yields and extra factor of $\ln x$ due to the $1/\zeta$ dependence of the splitting function.  The contribution to the RHS of \eqref{dglap+} at $\zeta\sim x$ for small $x$ arises from splittings which are not in the DGLAP kinematics. The main contribution for $\zeta\sim x \ll 1$ comes from splittings $(\frac{k^+}{\zeta}, \v q^2)\rightarrow (k^+,Q^2)+(\frac{k^+}{\zeta},\v q^2)$ with $Q^2\ll \v q^2\ll \frac{Q^2}{\zeta}$. These splittings are clearly in the BFKL kinematic region in terms of the longitudinal momentum distribution, and also outside of the double logarithmic region since the transverse momenta of the two daughter gluons are very different. Thus we find that for small $x$ the BFKL type splittings (beyond  the double logarithmic kinematics) give a nontrivial contribution to the DGLAP evolution via the dependence of the resolution scale on $\zeta$. How important is this contribution is an interesting open question.

\subsection{DGALP evolution -- nonlinear corrections}
Let us now consider the nonlinear corrections to the evolution of PDF stemming from the induced emission. 

\subsubsection{The virtual term}
First consider the nonlinear term in \eqref{css+}.
We write
\begin{equation}
\int_{\v k^2}^{Q^2} d^2\v p\mathcal{T}(\v p,x)\mathcal{T}(\v k,x)=\frac{1}{\pi}\frac{\partial}{\partial \v k^2}\left[\int_{0}^{Q^2} d^2\v p\mathcal{T}(\v p,x)\int_0^{\v k^2}d^2\v q\mathcal{T}(\v q,x)-\frac{1}{2}
\left[\int_0^{\v k^2}d^2\v q\mathcal{T}(\v q,x)\right]^2\right]
\end{equation}
Integrating this over $\v k$ up to $Q^2$ we have
\begin{equation}
\int^{Q^2} d^2\v k\int_{\v k^2}^{Q^2} d^2\v p\mathcal{T}(\v p,x)\mathcal{T}(\v k,x)=\frac{1}{2}\left[\int_{0}^{Q^2} d^2\v p\mathcal{T}(\v p,x)\right]^2=\frac{1}{2}G^2(Q^2,x)
\end{equation}
Thus the contribution to the evolution of PDF due to nonlinear term in \eqref{css+} is 
\begin{equation}\label{nl1}
\frac{\partial}{\partial \ln Q^2}[xG(Q^2,x)]_{virt}^{NL}=-\frac{\alpha_sN_c}{4\pi}\frac{1}{(2\pi)^3}\frac{1}{N_c^2-1}\frac{1}{Q^2S_\perp}[xG(Q^2,x)]^2
\end{equation}

\subsubsection{The real term}
In addition there is a contribution due to the real term \eqref{realtmddglap}. However, as we will argue now, the real term is effectively suppressed by an additional power of $\alpha_s$ relative to the virtual term. To understand the behavior of this term we first note that the integral over $\zeta$ in the nonlinear term is not likely to be dominated by small values of $\zeta\sim x$ even at small values of $x$, since the TMD on the RHS of \eqref{realtmddglap} then will have to be taken at very large longitudinal momentum. The frequency of gluons in this TMD is very small, and they have a lot of phase space to decay via the CSS mechanism discussed in the previous section. It therefore is likely that  the integral is dominated by the highest available value of $\zeta=1/2$. Let us for the sake of the argument approximate
\begin{equation}
\frac{1}{(1-\zeta)^{1/2}}x\frac{1-\zeta}{\zeta}\mathcal{T}(\v k, x\frac{1-\zeta}{\zeta})\approx \sqrt{2}x\mathcal{T}(\v k,x)
\end{equation}
The nonlinear term on the right hand side of \eqref{realtmddglap} then becomes (we assume the integral over $\eta$ between $E$ and $E+\Delta$ has been performed so that the $\delta$ function in \eqref{realtmddglap} has gone)
\begin{equation}
\frac{\alpha_sN_c}{\pi^2}\frac{\sqrt{2}}{16\pi^2}\frac{1}{Q^2S_\perp(N_c^2-1)}\int_x^{1/2}d\zeta\left[\frac{\zeta}{1-\zeta}+\frac{1-\zeta}{\zeta}+\zeta(1-\zeta)\right]\left[\frac{x}{\zeta}G(Q^2,\frac{x}{\zeta})\right]\left[x\mathcal{T}(\v k,x)\right]
\end{equation}
We can now use the linear approximation to  \eqref{realtmddglap} and \eqref{realtmdbfkl} to express $\mathcal{T}(\v k,x)$ in terms of the PDF. The resulting nonlinear contribution to the evolution of PDF then becomes
\begin{eqnarray}\label{nl2}
\frac{\partial}{\partial E}[xG(Q^2,x)]^{NL}_{real}&=&\frac{\alpha^2_sN_c}{\pi^3}\frac{\sqrt{2}}{16\pi^2}\frac{1}{Q^2S_\perp(N_c^2-1)}\int_x^{1/2}d\zeta\left[\frac{\zeta}{1-\zeta}+\frac{1-\zeta}{\zeta}+\zeta(1-\zeta)\right]\left[\frac{x}{\zeta}G(Q^2,\frac{x}{\zeta})\right]\nonumber\\
&\times&
\int_x^{1/2}d\xi\left[\frac{\xi}{1-\xi}+\frac{1-\xi}{\xi}+\xi(1-\xi)\right]\left[\frac{x}{\xi}G(\frac{Q^2}{\xi},\frac{x}{\xi})\right].
\end{eqnarray}
Thus indeed we get that  \eqref{nl2} is suppressed by an additional power of $\alpha_s$ relative to \eqref{nl1}.
Although we have explicitly showed this assuming the integral in \eqref{realtmddglap} is dominated by $\zeta=1/2$, this assumption is not important for the argument. The basic difference between the real and virtual contributions is that the virtual contribution to TMD has to be integrated over the transverse momentum in order to turn it into the contribution to DGLAP, while the real contribution is simply multiplied by the transverse resolution scale. The integral in the virtual piece turns the TMD in \eqref{css+} into PDF. While TMD at any fixed value of momentum (at least as long as the momentum is not in the nonperturbative domain) is at most of order $\alpha_s$, the integral over transverse momenta brings an extra logarithmic factor, so that the PDF is of order unity. However in the real term there is no integral over the transverse momentum, and the nonlinear correction to PDF is proportional to TMD at fixed transverse momentum. This is the origin of the extra factor of $\alpha_s$ in the real term as compared to the virtual contribution, and as such it does not depend on the value of $\zeta$ that dominates the integral over the longitudinal fraction.

 It follows therefore that at leading order the real contribution can be neglected and the entirety of the nonlinear correction comes from \eqref{nl1}. We caution again that for very small $x$ our dilute approximation should be modified, and the above conclusion may have to be revised. 
 
Interestingly the contribution \eqref{nl1} is negative, as it arises primarily from the virtual term. Therefore its effect is to slow down the DGLAP evolution, similarly to the effect of the GLR nonlinear correction, even though as discussed above,  the physics of \eqref{nl1} is very different from that of GLR, and parametrically the two corrections are different as well.

\section{Conclusions}
In this paper we applied the Born Oppenheimer evolution to observables of "partonic" type - gluon TMD and PDF. For this type of observables the frequency cutoff scale $E$ does not determine the energy of the process, but rather the resolution scale at which the observables are probed. The same cutoff $E$ determines both, the longitudinal and the transverse resolution scales.

In the context of BO approach we have shown that the evolution equation for TMD 
is nonlinear. The linear term in the equation
is equivalent to the CSS equation where the longitudinal and transverse resolution scales $\xi$ and $Q^2$, are related by the physical condition $\ln \frac{Q^2}{\v k^2}=\frac{1}{2}\ln \frac{k^+}{\xi}$. In this respect it is very different from the evolution based purely on decrease of the longitudinal momentum, where as shown in \cite{jimwlkcss}  the transverse resolution scale in the equivalent CSS equation is equal to the UV cutoff. We discussed in detail the correspondence between the BO cascade and the DGLAP/CSS  cascade and explained why the evolution they generate is identical (in the dilute limit) even though the cascades themselves are different in the sense that they populate different regions in the phase space. 

The equivalence between the BO evolution and the CSS evolution holds as long as $x\gg\xi$, which is indeed the regime where the CSS equation is valid. On the other hand the initial condition for this evolution, i.e. the value of the TMD at $x=\xi$ (the "real" contribution) that is generated by the BO cascade differs from the value in the standard perturbative calculation at very small $x\ll 1$. At such small values of $x$ the BO cascade contains contributions of splittings which are not in the usual DGLAP strongly ordered kinematics, but are rather generated by the BFKL splitting vertex.

The nonlinear term at large transverse momentum is suppressed by the factor $\frac{1}{N_c^2-1}\frac{1}{Q^2S_\perp}$ where $S_\perp\sim \frac{1}{\Lambda_{QCD}^2}$ is the transverse area of the projectile. It has a clear physical meaning of stimulated emission correction which enhances the probability of DGLAP splitting into a state with momentum $p$, if the unevolved state already contains particles with momentum $p$. 

Since this term is of the same order in $\alpha_s$ as the standard linear term, the suppression (disregarding the $1/N_c^2$ factor) is actually only strong at rather high transverse momentum. Only for gluons with $\v k^2>\frac{\Lambda_{QCD}^2}{\alpha_s}$ the induced emission term can be truly neglected, as for these momenta it is suppressed by a power of the coupling constant. Interestingly, the range of momenta $\Lambda_{QCD}<|\v k|<\frac{\Lambda_{QCD}}{g}$ was found to play a special role in the analysis of the density matrix of the CGC state in \cite{entanglement}. It was shown in \cite{entanglement} that in this semi hard momentum range gluon eigenmodes of the CGC density matrix behave as noninteracting massless two dimensional bosons. Not only that, but their contribution dominates the entanglement entropy of valence and soft gluons. It is interesting that the very same semi hard gluons exhibit stimulated emission in the context of BO evolution. Whether this is a coincidence or has some deeper meaning is a tantalizing question.

We obtain similar results for the evolution of the gluon PDF. We find that at not very small values of $x$ the linear part of the BO evolution of the PDF is equivalent to the DGLAP equation. This is true even though the glue-glue splitting function in the BO calculation is assembled from pieces that arise a little differently than in the DGLAP/CSS cascade. At very small $x$ the equation we obtain differs from the DGLAP due to the presence of non-DGLAP splittings alluded to earlier in the context of the real contribution to the TMD. The effect of these extra contributions is to modify the transverse resolution scale on the RHS of the DGLAP equation according to $Q^2\rightarrow Q^2/\zeta$. 

The nonlinear stimulated emission terms are also present in the BO equation for the PDF. These terms again are suppressed by the color factor $1/(N_c^2-1)$ as well as (at large $Q^2$) by the typical higher twist suppression factor $1/Q^2S_\perp$. 
%with $S$ being the transverse area of the projectile.  
Interestingly this correction is strongest for the virtual part of the evolution, for which it is of order $\alpha_s$ - i.e. is not suppressed by a power of $\alpha_s$ relative to the linear DGLAP term. For the real part the stimulated emission correction is of order $\alpha_s^2$, and is therefore suppressed by one extra power  of $\alpha_s$. As a result the net effect of the stimulated emission is to slow down the evolution of the gluon TMD. This slowdown mimics the effect of shadowing even though its physical origin is very different. 

We note that our analysis of the nonlinear effects was performed in the framework of the dilute approximation, discussed in detail in Section 2. We believe that this approximation is adequate at moderate values of $x$, but at small $x$ it may require serious modifications. Thus our conclusions about stimulated emission may not hold at low $x$.

It is worth stressing that the stimulated emission effect is very specific to the BO evolution. It arises in the BO approach due to the structure of the BO cascade. In the DGLAP/CSS cascade  partons with highest transverse momentum appear only in the 
very last step of the evolution irrespective of their longitudinal momentum. On the other hand in the BO cascade at any value of the evolution parameter $\ln k^-$ the wave function contains partons with high transverse momentum if their longitudinal momentum is high enough. Thus further evolution takes place in the presence of these high momentum partons and their nonvanishing density triggers stimulated emission.

The observable we considered in this paper are of "partonic" type, i.e. are defined at fixed longitudinal momentum which is not equal to the lowest longitudinal momentum present in the wave function. At moderate $x$ these observables are not strongly affected by low $x$ physics. This is not the type of observables that are of prime interest for example in BFKL or JIMWLK physics and not the ones that are expected to be strongly affected by saturation effects. In the latter context one is interested in gluons that most affect soft scattering. Those gluons do indeed have very small $x$. We will study the Born-Oppenheimer evolution of those observables in the next paper in the series \cite{three}.

  \section*{Acknowledgments}

 The research was supported by  the  Binational Science Foundation grant \#2012124;  MSCA RISE 823947 “Heavy ion collisions: collectivity and precision in saturation physics” (HIEIC), and by VATAT (Israel planning and budgeting committee) grant for supporting
theoretical high energy physics.
 The work of H.D. and A.K. is supported by the NSF Nuclear Theory grant \#2208387.
The work of ML was  funded by Binational Science Foundation grant \#2021789  and by the ISF grant \#910/23.\\
We  thank Physics Departments of the Ben-Gurion University and University of Connecticut for hospitality during mutual visits. \\
We also thank EIC Theory Institute, Brookhaven National Laboratory, CERN-TH group, 
National Centre for Nuclear Research, Warsaw;  Institute for Nuclear Theory at  the University of Washington, ECT$^*$,
 and ITP at the University of Heidelberg  for their support and hospitality during various stages of completing this project.

\bibliographystyle{unsrt}
\bibliography{BORGpaper2}

\end{document}